\documentclass[aps,pra,twocolumn,superscriptaddress,longbibliography]{revtex4}

\usepackage[english]{babel} 
\usepackage[latin1]{inputenc}
\usepackage[usenames,dvipsnames]{color}
\usepackage{subcaption}
\usepackage{graphicx}
\usepackage{amsmath,amssymb,amsfonts, physics}
\usepackage[colorlinks=true,linkcolor=blue,urlcolor=blue,citecolor=blue]{hyperref}
\usepackage[percent]{overpic}
\usepackage{comment}
\usepackage{float}

\begin{document}

\title{Charging Quantum Batteries via Dissipative Quenches}

\author{Riccardo Grazi}
\affiliation{Dipartimento di Fisica, Universit\`a di Genova, Via Dodecaneso 33, 16146, Genova, Italy}
\affiliation{CNR-SPIN, Via Dodecaneso 33, 16146, Genova, Italy}
\author{Donato Farina}
\affiliation{Dipartimento di Fisica ``Ettore Pancini'', Universit\`a degli Studi di Napoli Federico II, Complesso Universitario di Monte Sant\textquoteright Angelo, Via Cintia, 80126 Napoli, Italy}
\affiliation{Istituto Nazionale di
Fisica Nucleare, Sezione di Napoli, Via Cintia, 21,
Napoli, 80126, Italy}
\author{Niccol\'o Traverso Ziani}
\affiliation{Dipartimento di Fisica, Universit\`a di Genova, Via Dodecaneso 33, 16146, Genova, Italy}
\affiliation{CNR-SPIN, Via Dodecaneso 33, 16146, Genova, Italy}
\author{Dario Ferraro}
\affiliation{Dipartimento di Fisica, Universit\`a di Genova, Via Dodecaneso 33, 16146, Genova, Italy}
\affiliation{CNR-SPIN, Via Dodecaneso 33, 16146, Genova, Italy}

\begin{abstract}
We investigate work extraction in open quantum batteries composed of interacting spin chains weakly coupled to engineered environments. \textcolor{black}{Focusing primarily on two- and four-qubit XX models} initially prepared in thermal Gibbs states, we analyze how dissipation and dephasing, acting either locally or collectively, can generate and shape ergotropy during both transient and steady-state dynamics. By introducing \textcolor{black}{continuous interpolations between local and collective noise channels, as well as between amplitude damping and pure dephasing,} we systematically characterize the impact of environmental structure on work extractability. We show that purely dissipative dynamics can activate finite ergotropy from completely passive thermal states, giving rise to temperature-dependent transient regimes where hotter initial states temporarily outperform colder ones in an ergotropic Mpemba-like fashion. \textcolor{black}{We further derive the stationary ergotropy under local amplitude damping for arbitrary chain lengths and numerically demonstrate the persistence of the ergotropic Mpemba-like effect for chains of up to nine qubits.} In contrast, collective dissipation leads to steady states whose passivity crucially depends on the initial temperature and system size. \textcolor{black}{Finally, we show that local and collective dephasing lead to distinct behaviors: while local dephasing preserves the population of each total-excitation sector, collective dephasing leaves the initial Gibbs state invariant. Neither scenario gives rise to the transient ergotropic advantage observed under purely dissipative dynamics.}
\end{abstract}

\maketitle

\section{Introduction}

Since their conception~\cite{Alicki13}, quantum batteries (QBs) have emerged as a rapidly growing research area at the crossroad between quantum information and quantum thermodynamics~\cite{Bhattacharjee21, Quach23, Campaioli24, Camposeo25, Ferraro26}. They provide a versatile framework to explore fundamental aspects of energy storage and work extraction at the quantum scale, while also serving as a testing ground for quantum many-body physics. Beyond their foundational interest, QBs have been proposed as potential sources of coherent energy capable of powering more complex quantum devices, thereby establishing a direct link between quantum thermodynamics and quantum technologies~\cite{Chiribella21, Kurman25, Cioni26}. During the last decade, many models, as well as charging schemes, have been theoretically proposed as potential designs for QBs, including spin chains~\cite{Le18, Rossini19, Catalano24, Grazi24, Grazi25, farina2026charging}, harmonic oscillators~\cite{Andolina18, Hovhannisyan20, Cavaliere25, Andolina25, Cavaliere25b} and platforms for circuit quantum electrodynamics~\cite{Ferraro18, Ferraro19, Crescente20b, Carrasco22, Gemme23, Dou23, Rinaldi24, Rodriguez23, Erdman24, Canzio25, Massa25}, alongside some experimental realizations that have started to emerge~\cite{Quach22, Hu22, Gemme24, Joshi22, Cruz22, Tibben25, Hymas25, Hu26}. 

Even though the seminal works in the literature on energy storage and extraction have \textcolor{black}{mainly} focused on closed quantum systems as QBs~\cite{Alicki13, Campaioli24, Ferraro26, Binder15, Campaioli17}, in most realistic setups the role of an external environment cannot be neglected, shifting the attention towards the concept if open QBs (OQBs)~\cite{Farina19, Pirmoradian19, Zakavati21, Morrone23, Crotti26}. In the simplest scenario, the system acting as a QB weakly interacts with a memoryless (Markovian) environment, allowing one to investigate the effects of dissipation and dephasing on the performance of the device. In this regime, the time evolution of the QB matrix $\rho(t)$ can be described by a master equation in Gorini--Kossakowski--Sudarshan--Lindblad (GKSL) form~\cite{Petruccione_book},
\begin{equation}
\dot{\rho}(t)
= -i[H,\rho(t)]
+ \mathcal{D}[\rho],
\label{time_local_master}
\end{equation}
where $H$ is the system Hamiltonian and
\begin{equation}
    \mathcal{D}[\rho] = \sum_i \gamma_i
\left(
L_i \rho(t) L_i^\dagger
-\frac{1}{2}\{L_i^\dagger L_i,\,\rho(t)\}
\right)
\end{equation}
is usually known as the Lindblad superoperator, where $L_i$ are jump operators and $\gamma_i$ are non-negative damping rates. In the absence of environmental effects, i.e., when all $\gamma_i=0$, Eq.~\eqref{time_local_master} reduces to the conventional von Neumann equation for the time evolution of a density matrix.

Once the system and the environment are specified, the structure of the jump operators can, in principle, be derived from a microscopic description of their mutual interaction. These operators encode the dissipative part of the dynamics and characterize how the environment acts on the system degrees of freedom. When the system under investigation is a many-body QB, two types of interactions coexist: interactions among the subsystems and interactions between each subsystem and the environment. If these energy scales are comparable, a consistent description generally requires adopting the so-called global approach~\cite{Rivas10, Hofer17, Cattaneo2019, Farina20}, in which the jump operators are derived microscopically from the full system--environment Hamiltonian. Conversely, if the interactions among subsystems are sufficiently weak, it is sometimes possible to employ the local approach~\cite{Rivas10, Hofer17, Cattaneo2019, Farina20}. Within this approximation, each subsystem is treated as independently coupled to the environment and the dissipative dynamics is described in terms of local jump operators. Although this approach is often simpler and widely used, it may lead to unphysical predictions if applied outside its regime of validity. Nevertheless, it has been shown that suitably engineered dissipation channels can render the local approach a meaningful and accurate effective description in several physical settings, raising the question of when and under which conditions such an approximation can be justified.

Regardless of this distinction, several works have already shown that dissipation can \textcolor{black}{have a favorable impact at the level of the charging dynamics of OQBs \cite{Barra19, Tian25}. This is not limited to the conventional Lindblad approximation~\cite{ Tabesh20, Cakmak2020, Ghosh21, Ahmadi24, Choquehuanca2024, Ahmadi25, Oularabi25}, but has been observed also within the Caldeira-Leggett~\cite{Hovhannisyan20,Feliu24, Bhanja24, Cavaliere25, Cavaliere25b} and collisional~\cite{Mayo22} descriptions, where strong system-environment coupling and memory effects can be taken into account~\cite{Morrone23, Massa26}. In particular, environment-induced dissipation can positively affect} relevant figures of merit such as ergotropy, which quantifies the maximum amount of work that can be extracted from a quantum state by means of unitary operations ~\cite{Allahverdyan04}
\begin{equation}
\mathcal{E}(\rho,H) = \Tr(\rho H) - \min_{U}\Tr\!\left(U\rho U^\dagger H\right),
\end{equation}
where the minimization is performed over all unitary operators $U$. Remarkably, it has been demonstrated that dissipative dynamics can activate finite ergotropy even when the system is initially prepared in a passive state, from which no work can be extracted \cite{Hadipour2025}. These findings highlight the importance of understanding how environmental effects, rather than being purely detrimental, can be harnessed as a resource. 

\textcolor{black}{To address this problem for simple and analytically treatable cases,} here we \textcolor{black}{primarily} investigate two- and four-qubit systems modeled as XX spin-chain OQBs, initially prepared in thermal Gibbs states, which are by definition completely passive, i.e., passive for any number of copies of the system ~\cite{Pusz78}. With the aim of analyzing the resulting dynamics in terms of work extractability, we consider a weak coupling between the OQB and external baths by engineering dissipative and dephasing channels that can act either locally on each qubit or collectively on the entire system. \textcolor{black}{While the main analysis focuses on these two system sizes, we extend selected results to longer chains, both analytically and numerically. In particular, we characterize the stationary ergotropy under local amplitude damping for arbitrary chain lengths and investigate the persistence of the ergotropic Mpemba-like effect for systems containing up to nine qubits.}

More specifically, we introduce a continuous interpolation between local and global dissipative schemes, as well as between pure dissipation and pure dephasing channels. This allows us to explore how the structure of the environment affects the charging and relaxation properties of the QB. Within this framework, we study the ergotropy both during the transient dynamics and in the non-equilibrium steady state, highlighting how the initial temperature of the thermal state influences the generation and evolution of ergotropy throughout the entire dynamics, providing a systematic characterization of temperature-dependent effects in OQBs.

We demonstrate that
purely dissipative dynamics can activate ergotropy,
giving rise to temperature-dependent transient regimes where hotter initial states temporarily outperform
colder ones in an ergotropic Mpemba-like fashion. 
In this regard, we notice that while \textcolor{black}{an analogous phenomenology} have been explored in the context of energy and ergotropy storage, \textcolor{black}{where the form of the initial states and the non-monotonic evolution of the passive energy in time seem crucial \cite{Medina25, sapui2026ergotropic}, its manifestation at the level of} ergotropy generation via dissipative charging remains largely unexplored. Besides, we find that collective dissipation leads
to steady states whose passivity crucially depends on the initial temperature and system size. Finally, we demonstrate
that \textcolor{black}{local and collective dephasing lead to qualitatively different behaviors: while local dephasing preserves the total number of excitations, collective dephasing leaves the initial Gibbs state invariant. In neither case we observe transient ergotropic advantages for hotter initial states.}

The manuscript is organized as follows. In Sec. II, we introduce the general formalism of the dissipative charging scheme and present the specific models considered. In Sec. III, we report our results for the different dissipative charging protocols. Section IV is devoted to the analysis of purely dephasing environments. Finally, in Sec. V, we summarize our conclusions and discuss perspectives for future work. \textcolor{black}{Five} Appendices are devoted to technical details of the calculations and additional discussions. 

\section{GENERAL FORMALISM AND CONCRETE IMPLEMENTATIONS} 

\begin{figure}
\centering
\includegraphics[width=.85\linewidth]{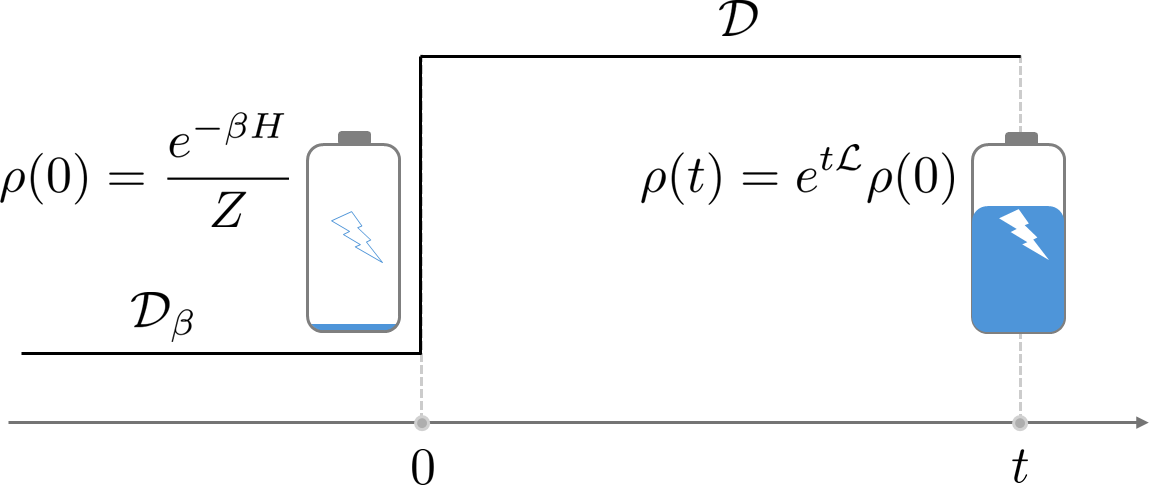}
\caption{Scheme of the considered dissipative quench protocol. The system is initialized in the Gibbs state in Eq. \eqref{input}. From time $0$ to time $t$ it evolves according to a nontrivial open system dynamics obtained via suddenly changing the dissipator. This allows one to charge the system in ergotropy uniquely via dissipative dynamics.}
\label{fig:scheme-charging}
\end{figure}

Throughout this work, we consider as initial "discharged" state 
the Gibbs state 
\begin{equation}
\rho(0)=\rho_{\beta}\equiv\frac{e^{-\beta H}}{{\rm Tr}(e^{-\beta H})}
\label{input}
\end{equation}
associated to the system Hamiltonian $H$, \textcolor{black}{with} $\beta$ being the inverse temperature.
We can assume, for instance, such Gibbs state to be naturally prepared as the steady state of the well known Davies  
generator satisfying Kubo-Martin-Schwinger (KMS) condition (see, e.g. \cite{Guo2025designingopen} and references therein), that we indicate as 
\begin{equation}
\mathcal{L}_\beta\, \bullet=-i[H, \bullet]+\mathcal{D}_\beta\, \bullet ,
\end{equation}
with
$\mathcal{L}_\beta(\rho_\beta)=0$.
At time $t=0$ we quench only the dissipative part of the Lindbladian, suddenly switching it from $\mathcal{D}_\beta$ to a new dissipator $\mathcal{D}$.
The new Lindbladian reads,
\begin{equation}
\mathcal{L}\, \bullet=-i[H, \bullet]+\mathcal{D}\, \bullet\,,
\end{equation}
and \textcolor{black}{leads to the evolution of the state density matrix in time according to} 
$\rho(t)=e^{ \mathcal{L} t}\rho(0)$.
Formally, \textcolor{black}{this is achieved via the} step-like Lindbladian,
\begin{equation}
-i[H,\bullet]+\theta(-t) \mathcal{D}_\beta \,\bullet+
\theta(t) \mathcal{D} \,\bullet\,, \end{equation}
\textcolor{black}{realizing} the dissipative charging protocol schematically depicted in Fig.\,\ref{fig:scheme-charging}.
In this regard, a necessary condition for the evolution to be nontrivial is $\mathcal{L}\,\rho_\beta\neq 0$.
Concrete implementations able to create ergotopy via dissipation are now introduced.

Our OQB consists of a spin chain described by the XX Hamiltonian
\begin{equation}
    H = J\sum_{i = 1}^{N-1}(\sigma_i^x \sigma_{i+1}^x + \sigma_i^y \sigma_{i+1}^y) +h \sum_{i = 1}^{N}\sigma_i^z \label{Hamiltonian_XX}
\end{equation}
where $J$ denotes the overall scale of energy (from now on, $J = 1$), $h$ is the external transverse field, $\sigma_i^\alpha$ (with $\alpha = x,y,z$) are the conventional Pauli
matrices corresponding to the $i$-th spin and $N$ is the number of qubits. In this work, we \textcolor{black}{mainly} focus on systems composed of $N=2$ and $N=4$ qubits. The case $N=2$ represents the simplest nontrivial QB, being the minimal setting in which correlations between subsystems can emerge, while still allowing for partial analytical insight into the dynamics~\cite{Razzoli25}. The choice of $N=4$ is motivated by the fact that some of the qualitative features discussed in this work appear for larger system sizes, already for $N=3$, but become more pronounced and clearly identifiable starting from four qubits. 

\textcolor{black}{The effects of the interaction with the external environment are encoded into} two Lindblad superoperators. The first one corresponds to the dissipative channel and its action on the density matrix reads
\begin{equation} \label{Superoperator_Sigma_-}
    \mathcal{D}^{(-)} [\rho] = \sum_{i,j} \Gamma_{ij}^{(-)} (\sigma_i^- \rho \sigma_j^+ - \frac{1}{2}\{\sigma_j^+\sigma_i^-,\rho\}),
\end{equation}
with
\begin{equation} \label{Gamma^-}
     \Gamma_{ij}^{(-)} = \gamma\left[(1 - \alpha^{(-)}) \delta_{ij} + \alpha^{(-)}\right]
\end{equation}
where $\alpha^{(-)}$ is an interpolation parameter in the range $[0,1]$, ensuring complete positivity of the dynamics. In the limit $\alpha^{(-)}=0$, the system is in the local (parallel) dissipation regime, yielding
$\Gamma_{ij}^{(-)}=\gamma\delta_{ij}$, so that each qubit dissipates independently at the same rate.
In this case, the superoperator in Eq.~\eqref{Superoperator_Sigma_-} is equivalent to a set of $N$ local jump operators \textcolor{black}{of the form}
\begin{equation} \label{Jump_Op_Local}
    L_i = \sqrt{\gamma}\,\sigma_i^-, \qquad i=1,\dots,N.
\end{equation}
This dissipation model admits a microscopic interpretation in terms of a cavity-mediated interaction among the spins composing the QB. In particular, the single-qubit jump operator in Eq.~\eqref{Jump_Op_Local} naturally emerges from a Jaynes-Cummings-type coupling between the qubit and a bosonic mode, in the regime where the cavity dynamics is much faster than the qubit evolution~\cite{Schleich2015}. In this limit, the photon emitted by the qubit leaves the cavity before it can be reabsorbed, so that no coherent energy exchange can build up between the two systems. As a consequence, the cavity acts as an effective Markovian reservoir and its dynamics can be adiabatically eliminated, leading to a purely dissipative evolution for the qubit. A detailed derivation of this mapping is presented in Appendix \ref{App:JC}.

In the opposite limit $\alpha^{(-)}=1$, the dissipation becomes fully collective, with
$\Gamma_{ij}^{(-)}=\gamma$ for all sites. This corresponds to a single collective jump operator acting on the entire system,
\begin{equation} \label{Jump_Op_Collective}
    L = \sqrt{\gamma}\sum_{i=1}^N \sigma_i^-.
\end{equation}
In analogy with the mapping for local dissipation, the collective jump operator in Eq.~\eqref{Jump_Op_Collective} can be derived by considering $N$ identical qubits coupled to a common cavity mode described by the Tavis-Cummings Hamiltonian~\cite{Kirton19}, as discussed in Appendix \ref{App:JC}.

The second Lindblad superoperator we introduce defines the dephasing channel, such that
\begin{equation}
    \mathcal{D}^{(z)} [\rho] = \sum_{i,j} \Gamma_{ij}^{(z)} (\sigma_i^z \rho \sigma_j^z - \frac{1}{2}\{\sigma_j^z\sigma_i^z,\rho\}),
\end{equation}
with
\begin{equation}
    \Gamma_{ij}^{(z)} = \gamma\left[(1 - \alpha^{(z)}) \delta_{ij} + \alpha^{(z)}\right].
\end{equation}
Similarly to the dissipation channel, interpolating $\alpha^{(z)}$ results in mixed configurations between local ($\alpha^{(z)} = 0$) and collective ($\alpha^{(z)} = 1$) schemes, with jump operators analogous to the ones reported in Eqs. \eqref{Jump_Op_Local} and \eqref{Jump_Op_Collective} replacing $\sigma_i^-$ with $\sigma_i^z$. 

Once the \textcolor{black}{dissipative} channels are defined, we aim at studying the behavior of two- and four-qubit systems described by the Hamiltonian of Eq. \eqref{Hamiltonian_XX} evolving according to the Lindblad master equation of Eq. \eqref{time_local_master}, where
\begin{equation} \label{Full_Superoperator}
    \mathcal{D}[\rho] = (1 - \alpha)\mathcal{D}^{(-)}[\rho] + \alpha \mathcal{D}^{(z)}[\rho].
\end{equation}
In the main text we report the single-channel cases $\alpha = 0$ (Sec. \ref{Sec:alpha_zero}) and $\alpha = 1$ (Sec. \ref{Sec:alpha_one}), while in Appendix \ref{App:Interpol} plots for \textcolor{black}{intermediate values of} $\alpha \in (0,1)$ are shown.


\section{Results on Dissipative Charging} \label{Sec:alpha_zero}
\subsection{Results for arbitrary chain length}
\begin{figure}
    \centering
    \includegraphics[width=0.9\linewidth]{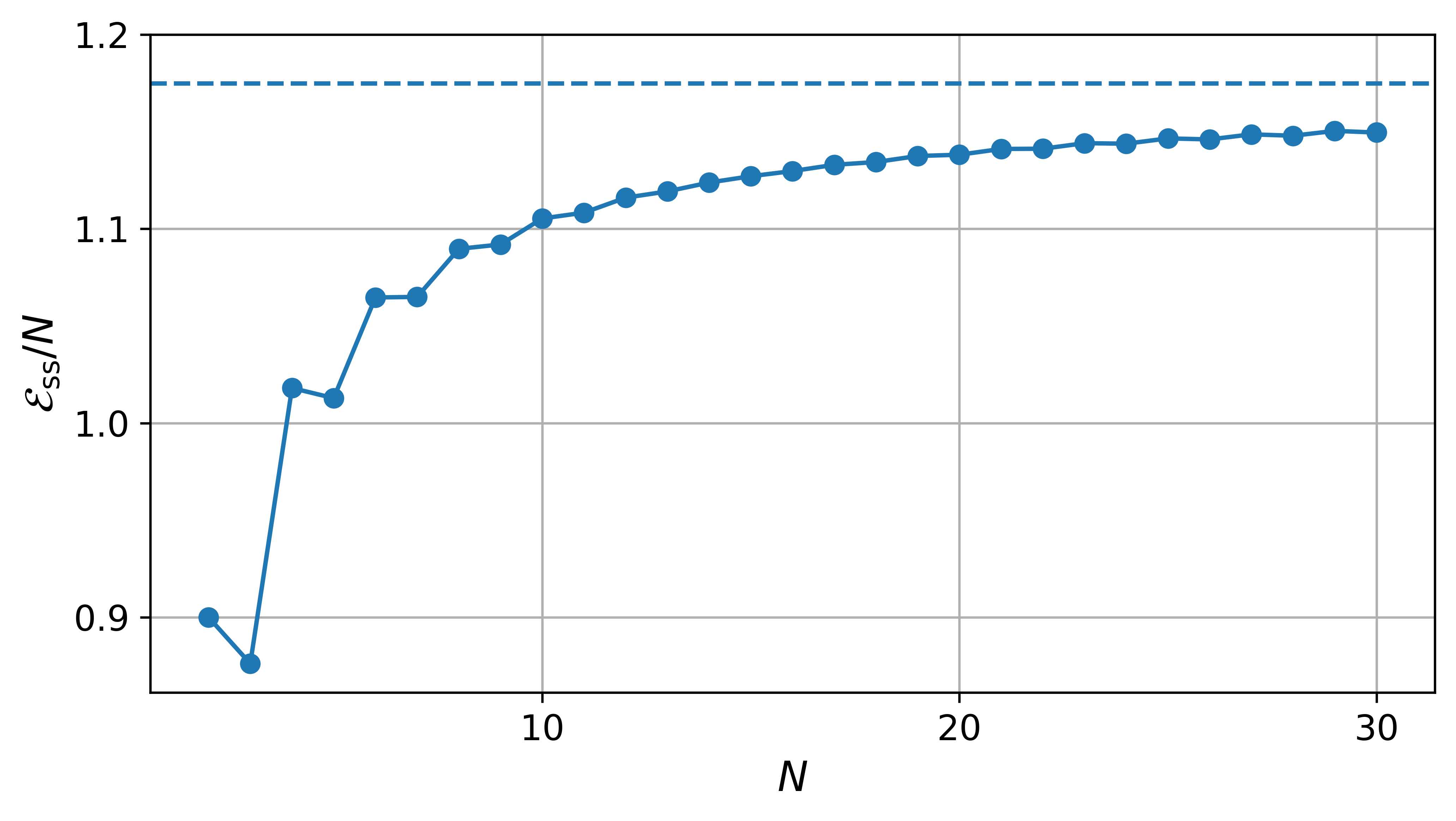}
    \caption{\textcolor{black}{Stationary ergotropy density $\mathcal{E}_{SS}/N$ as a function of the size $N$ of the XX spin chain for $J = 1$ and $h = 0.1$. The dashed line corresponds to the asymptotic value \(1.17483\) obtained in the thermodynamic limit for our set of parameters.}}
    \label{E_SS_Locale_N_Generico}
\end{figure}

We start by analyzing the parallel dissipation regime obtained by fixing $\alpha^{(-)}=0$ in Eq. \eqref{Gamma^-}. \textcolor{black}{Before focusing on the dynamics of the simplest two-qubit system, we can infer a general property of the stationary state for an arbitrary chain length $N$ from the structure of the jump operators in Eq. \eqref{Jump_Op_Local}. Indeed, local amplitude damping drives each spin towards its ground state, so that the unique stationary state is the fully polarized state $\ket{g}^{\otimes N}$, independently of the initial state. This allows us to obtain an analytical expression for the stationary ergotropy, which is given by the energy difference between this state and the ground state of the Hamiltonian,
\begin{equation}
\mathcal{E}_{\mathrm{SS}}=-Nh-E_0(N),
\end{equation}
where \(E_0(N)\) is the ground-state energy. By mapping the XX Hamiltonian of Eq. \eqref{Hamiltonian_XX} onto free fermions \cite{Franchini17}, the single-particle energies are found to be
\begin{equation}
\epsilon_m=2h+4J\cos\left(\frac{m\pi}{N+1}\right),
\qquad m=1,\ldots,N.
\end{equation}
Therefore, the ground-state energy is obtained by filling all modes with negative energy, yielding
\begin{equation}
\mathcal{E}_{\mathrm{SS}}=-\sum_{\epsilon_m<0}\epsilon_m.
\end{equation}
In the thermodynamic limit, for \(0<h<2J\), this gives
\begin{equation}
\lim_{N\to\infty}\frac{\mathcal{E}_{\mathrm{SS}}}{N}
=\frac{1}{\pi}\left[
4J\sqrt{1-\frac{h^2}{4J^2}}
-2h\arccos\left(\frac{h}{2J}\right)
\right].
\end{equation}
For the parameters we consider in this work, i.e. $J = 1$ and $h = 0.1$, the resulting steady-state ergotropy density is approximately \(1.17483\). This result shows that the stationary ergotropy is extensive and independent of the initial temperature, as shown in Figure \ref{E_SS_Locale_N_Generico}.}
\subsection{Two-qubit system} \label{Subsec:alpha_zero_two_qubits}
\begin{figure}[H]
    \centering
    \includegraphics[width=\linewidth]{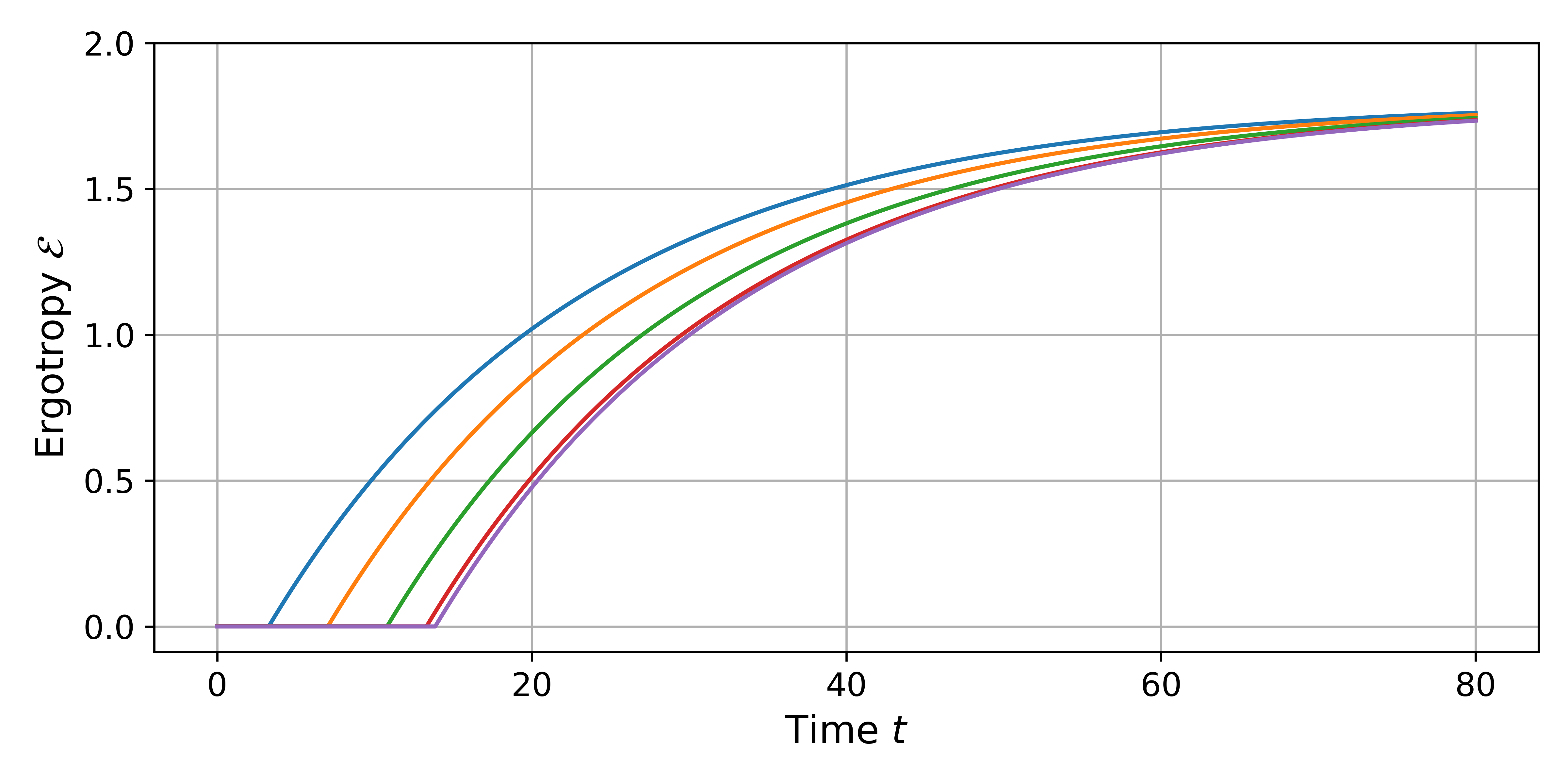}
    \caption{Ergotropy $\mathcal{E}$ of the two-qubit system as function of time and in presence of two parallel dissipative channels for $h = 0.1$, $\gamma = 0.05$ and initial state's temperature $\beta = 0.2$ (blue), $\beta = 0.5$ (orange), $\beta = 1$ (green), $\beta = 2$ (red) and $\beta = 5$ (purple). Local dissipation activates ergotropy in the transient regime at different times $t_c$ \textcolor{black}{according to} Eq. \eqref{t_c}, while all temperatures asymptotically converge to the same non-passive steady state \textcolor{black}{characterized by the ergotropy derived in Eq. (\ref{Epsilon_inf})}.}    \label{fig:ergotropy_diss_parallel}
\end{figure}

\textcolor{black}{After this general discussion, we can now focus on the two-qubit QB.} In the full-dissipation scenario, both the Hamiltonian and the jump operators can only preserve or decrease the total number of excitations, so that the full set of coupled equations for the elements of the density matrix $\rho$ breaks up into three disjoint sectors: a two-excitation sector $\{\rho_{ee,ee}\}$, a one-excitation sector $\{\rho_{eg,eg},\rho_{ge,ge},\rho_{eg,ge},\rho_{ge,eg}\}$, which is the only nontrivial block where dissipation leaks probability downward, and a zero-excitation sector $\{\rho_{gg,gg}\}$, which is fed by the one-excitation manifold but does not feed back. Within this scheme, starting from the Lindblad master equation, we obtain the following set of equations
\begin{equation}
    \begin{cases}
        &\dot{p}_{gg} = \gamma (p_{eg} + p_{ge}) \\
        &\dot{p}_{eg} = -2i(c^* - c) - \gamma p_{eg} + \gamma p_{ee}\\
        &\dot{p}_{ge} = -2i(c - c^*) - \gamma p_{ge} + \gamma p_{ee}\\
        &\dot{p}_{ee} = -2\gamma p_{ee} \\
        &\dot{c} = -2i(p_{ge} - p_{eg}) - \gamma c
    \end{cases} \label{ODEs_twoqubits_parallel}
\end{equation}
where we have introduced the short notation 
\begin{equation*}
    p_{nm} \equiv \rho_{nm,nm}, \quad \quad c \equiv \rho_{eg,ge}.
\end{equation*}
After solving the above equations, the ergotropy for different values of the initial state's inverse temperature $\beta$ can be plotted, as shown in Fig. \ref{fig:ergotropy_diss_parallel}. \textcolor{black}{We} observe that the time $t_c$ at which \textcolor{black}{a finite ergotropy starts to emerge} depends on temperature; in particular, the hotter the initial state is, the sooner ergotropy becomes non-zero. It is possible to show that $t_c$ corresponds to the time at which pairs of eigenvalues of $\rho(t)$ cross\textcolor{black}{, thereby changing the ordering of the populations with respect to the energy eigenstates. This condition can be computed analytically starting from the solutions of} Eqs. \eqref{ODEs_twoqubits_parallel} and it's given by
\begin{equation} \label{t_c}
    t_c = \frac{1}{\gamma} \ln[1 + \tanh(\beta - \beta  h)].
\end{equation}
Importantly, these crossings appear only if the condition $h < J$ is satisfied. \textcolor{black}{The physical origin of this condition can be understood by studying the energy spectrum of the two-qubit Hamiltonian. In particular, the singlet state \(\ket{\psi^-}=(\ket{eg}-\ket{ge})/\sqrt{2}\) and the fully polarized state \(\ket{gg}\) have energies \(E_-=-2J\) and \(E_{gg}=-2h\), respectively. These two levels cross at \(h=J\), marking a change in the ground state of the system. For \(h<J\), the singlet is the ground state, while local amplitude damping, as previously discussed, drives the system towards \(\ket{gg}\). The resulting steady state is therefore non-passive and independent of temperature, retaining a finite ergotropy given by
\begin{equation}
    \mathcal{E}_{SS} \equiv \mathcal{E}(\infty) = \langle gg|H|gg\rangle - \langle \psi^-|H|\psi^-\rangle 
= 2(1 - h),
\label{Epsilon_inf}
\end{equation}
which yields $\mathcal{E}_{SS} = 1.8$ for the value $h=0.1$ used in Fig. \ref{fig:ergotropy_diss_parallel}.
On the other hand, for \(h\geq J\), \(\ket{gg}\) is itself a ground state of the Hamiltonian, and the stationary ergotropy vanishes. So, local dissipation is able to generate non-zero extractable work when its asymptotic target is not an energy-minimizing state.} We can also notice that Eq. \eqref{t_c} is asymptotically bounded in terms of $\beta$, since, as $\beta$ goes to infinity \textcolor{black}{(zero temperature limit)}, $t_c$ approaches $\ln(2)/\gamma$ \textcolor{black}{($t_c\approx 13.86$ for the value $\gamma=0.05$ used in Fig. \ref{fig:ergotropy_diss_parallel}). This upper bound can be explained by observing that, in the zero-temperature limit, the weight of the initially occupied state decays as $e^{-\gamma t}$, while the population transferred to $|gg\rangle$ grows correspondingly as $1-e^{-\gamma t}$. These results can be easily obtained by solving the system in Eq. \eqref{ODEs_twoqubits_parallel} in the $\beta \to \infty$ limit. So, ergotropy is activated when these two contributions become equal, which occurs exactly after a time $\ln(2)/\gamma$. At finite temperatures, however, the occupation of higher-energy states modifies this balance, accounting for the temperature dependence of $t_c$.}

\begin{figure}[H]
    \centering
    \includegraphics[width=\linewidth]{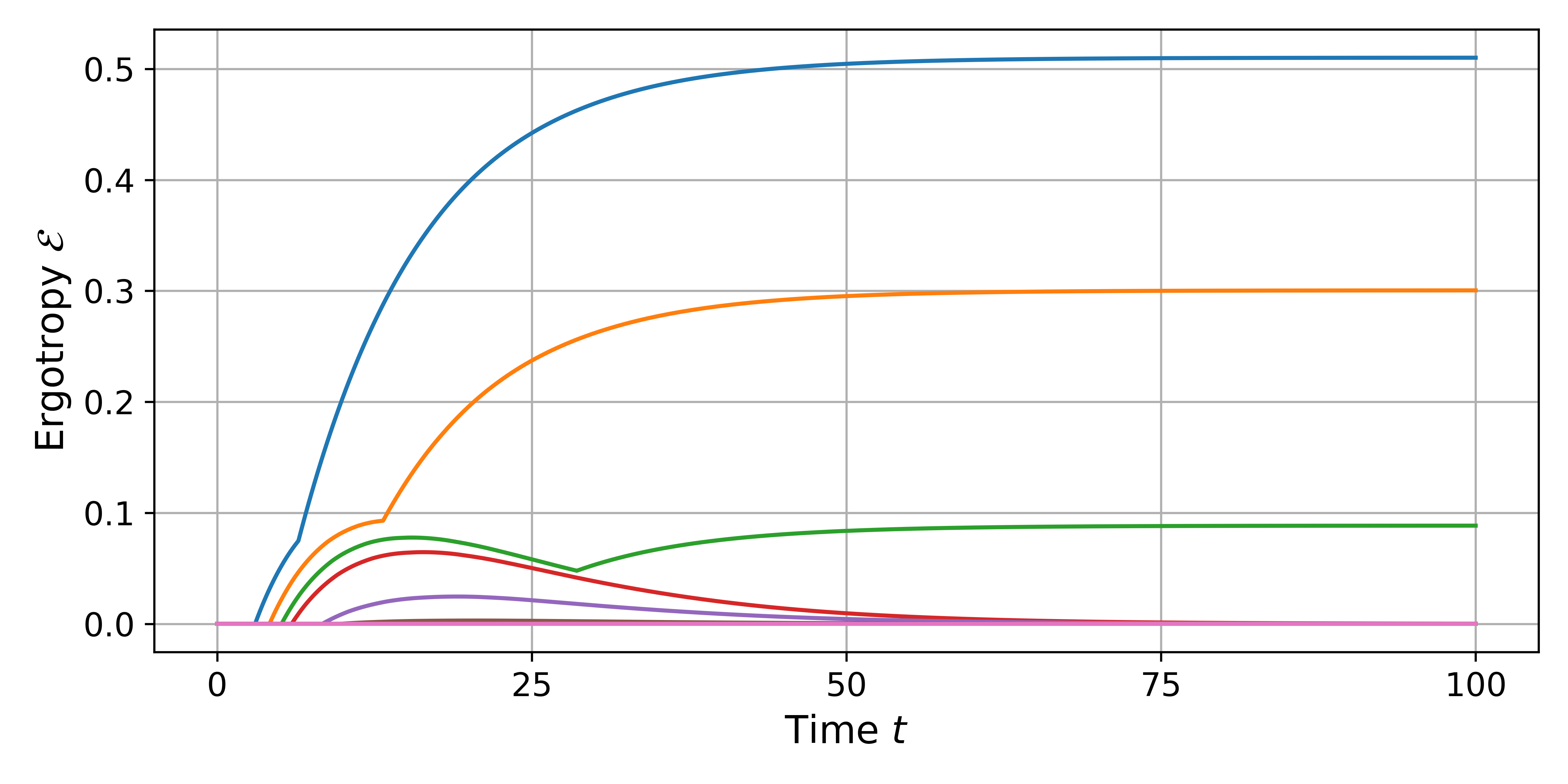}
    \caption{Ergotropy $\mathcal{E}$ of the two-qubit system as function of time in presence of a collective dissipative channel for $h = 0.1$, $\gamma = 0.05$ and initial state's temperature $\beta = 0.2$ (blue), $\beta = 0.3$ (orange), $\beta = 0.4$ (green), $\beta = 0.5$ (red), $\beta = 1$ (purple), $\beta = 2$ (brown) and $\beta = 5$ (pink). Collective dissipation leads to a temperature-dependent steady state, with a critical inverse temperature \textcolor{black}{$\beta_{c}\approx 0.44$} separating passive and non-passive regimes.}
    \label{fig:ergotropy_diss_collective}
\end{figure}

\begin{figure}[H]
    \centering
    \includegraphics[width=\linewidth]{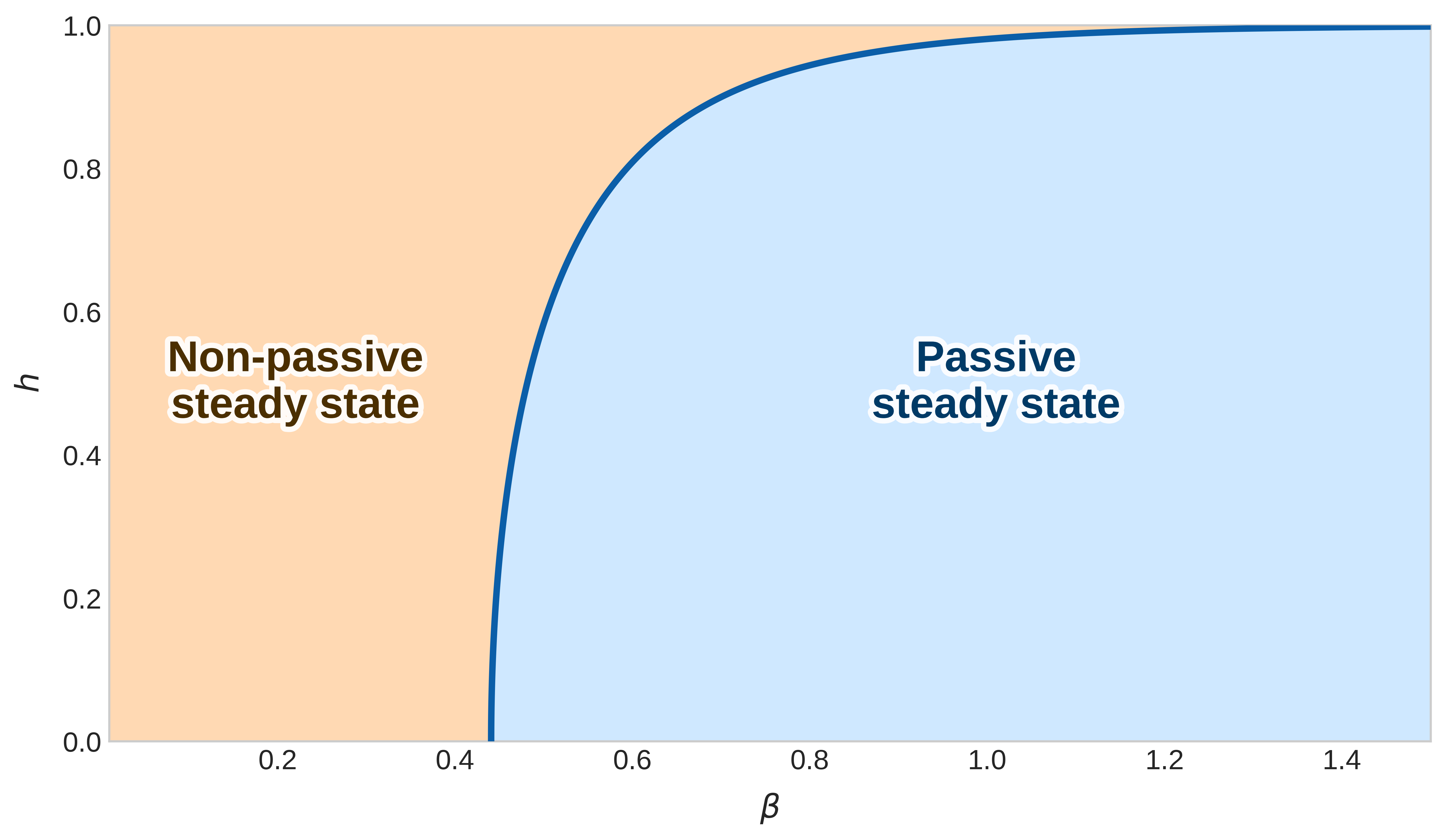}
    \caption{Phase diagram of the system's steady state. The non-passive and passive regions are separated by the blue thick curve representing the critical value $\beta_c$ as function of $h$, obtained by numerically solving Eq. \eqref{Passive_Non_Passive_Boundary}.}
    \label{fig:phase_diagram}
\end{figure}

\noindent Now we focus on the collective case $\alpha^{(-)}=1$. The analog of Eqs. \eqref{ODEs_twoqubits_parallel} for this new scenario is
\begin{equation}
    \begin{cases}
        &\dot{p}_{gg} = \gamma (p_{eg} + p_{ge}) + \gamma(c + c^*) \\
        &\dot{p}_{eg} = -2i(c^* - c)- \gamma p_{eg} + \gamma p_{ee}- \frac{\gamma}{2}(c + c^*)\\
        &\dot{p}_{ge} = -2i(c - c^*) - \gamma p_{ge} + \gamma p_{ee} - \frac{\gamma}{2}(c + c^*)\\
        &\dot{p}_{ee} = -2\gamma p_{ee} \\
        &\dot{c} = -2i(p_{ge} - p_{eg})- \gamma c + \gamma p_{ee} - \frac{\gamma}{2}(p_{eg} + p_{ge}).
    \end{cases} \label{ODEs_twoqubits_collective}
\end{equation}
It is important to highlight that, when we consider a collective bath, coherences influence populations, as shown by Eqs. \eqref{ODEs_twoqubits_collective}. When the dynamical evolution of ergotropy is plotted \textcolor{black}{(see Fig. \ref{fig:ergotropy_diss_collective})} it is possible to distinguish two classes of curves: if the initial state's inverse temperature is below a certain \textcolor{black}{critical} value $\beta_{c}$ the ergotropy reaches a non-passive steady state whose value depends on the initial temperature itself, while above $\beta_{c}$ ergotropy drops to zero, reaching a passive steady state. \textcolor{black}{Considering the long-time behavior of density matrix's eigenvalues, it is possible to show (see Appendix \ref{App:beta_c} for the detailed derivation)} that such critical value of the inverse temperature $\beta_c$ satisfies the equation
\begin{equation} \label{Passive_Non_Passive_Boundary}
    \sinh(2\beta_c) = \cosh(2\beta_c h).
\end{equation}
For the situation represented in Fig. \ref{fig:ergotropy_diss_collective} \textcolor{black}{($h=0.1$)}, a numerical resolution of Eq. \eqref{Passive_Non_Passive_Boundary} leads to $\beta_c \approx 0.44$, in full agreement with the observed behavior. \textcolor{black}{For $\beta < \beta_c$, the stationary ergotropy can be explicitly written as
\begin{equation}
    \mathcal{E}_{\mathrm{SS}}=2(1-h)(1-2s_\infty),
\end{equation}
with
\begin{equation}
    s_\infty=\frac{e^{2\beta}}{2[\cosh(2\beta)+\cosh(2\beta h)]}.
\end{equation}
}
\noindent Fig. \ref{fig:phase_diagram} illustrates the emergence of these two qualitatively distinct steady-state regimes in the $\beta$-$h$ plane. The \textcolor{black}{blue} boundary curve, given by Eq. \eqref{Passive_Non_Passive_Boundary}, separates the non-passive region (\textcolor{black}{orange region} on the left), where the steady state retains finite ergotropy and thus allows for work extraction, from the passive region (\textcolor{black}{light blue region} on the right), where the steady state is fully thermal and no extractable work remains.

\subsection{Four-qubit system}
\begin{figure}[H]
  \centering

  \begin{subfigure}{\columnwidth}
    \centering
    \includegraphics[width=\columnwidth]{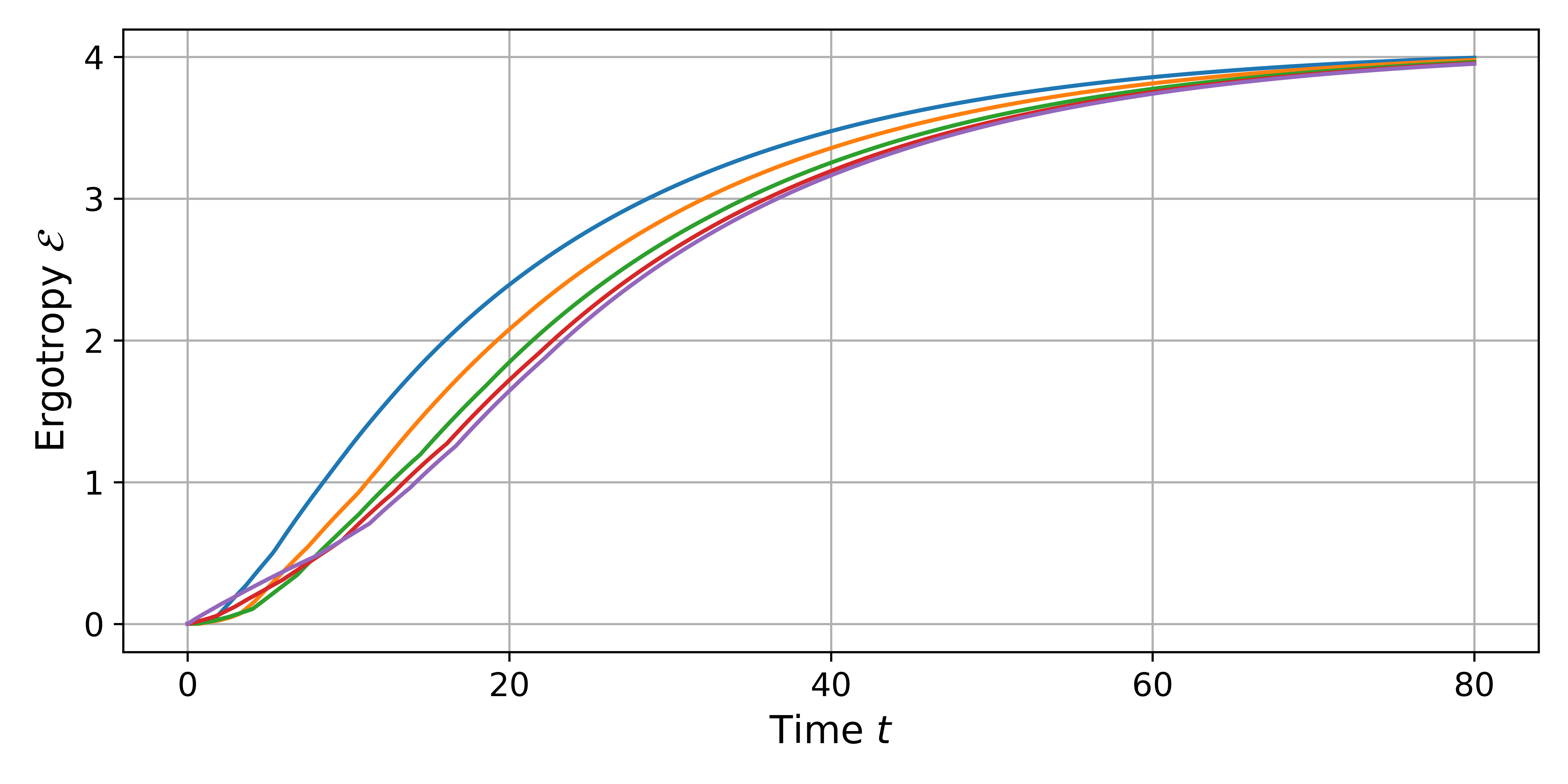}
    \caption{}
    \label{Four_Qubits_Ergotropies_Parallel}
  \end{subfigure}

  \vspace{0.4cm}

  \begin{subfigure}{\columnwidth}
    \centering
    \includegraphics[width=\columnwidth]{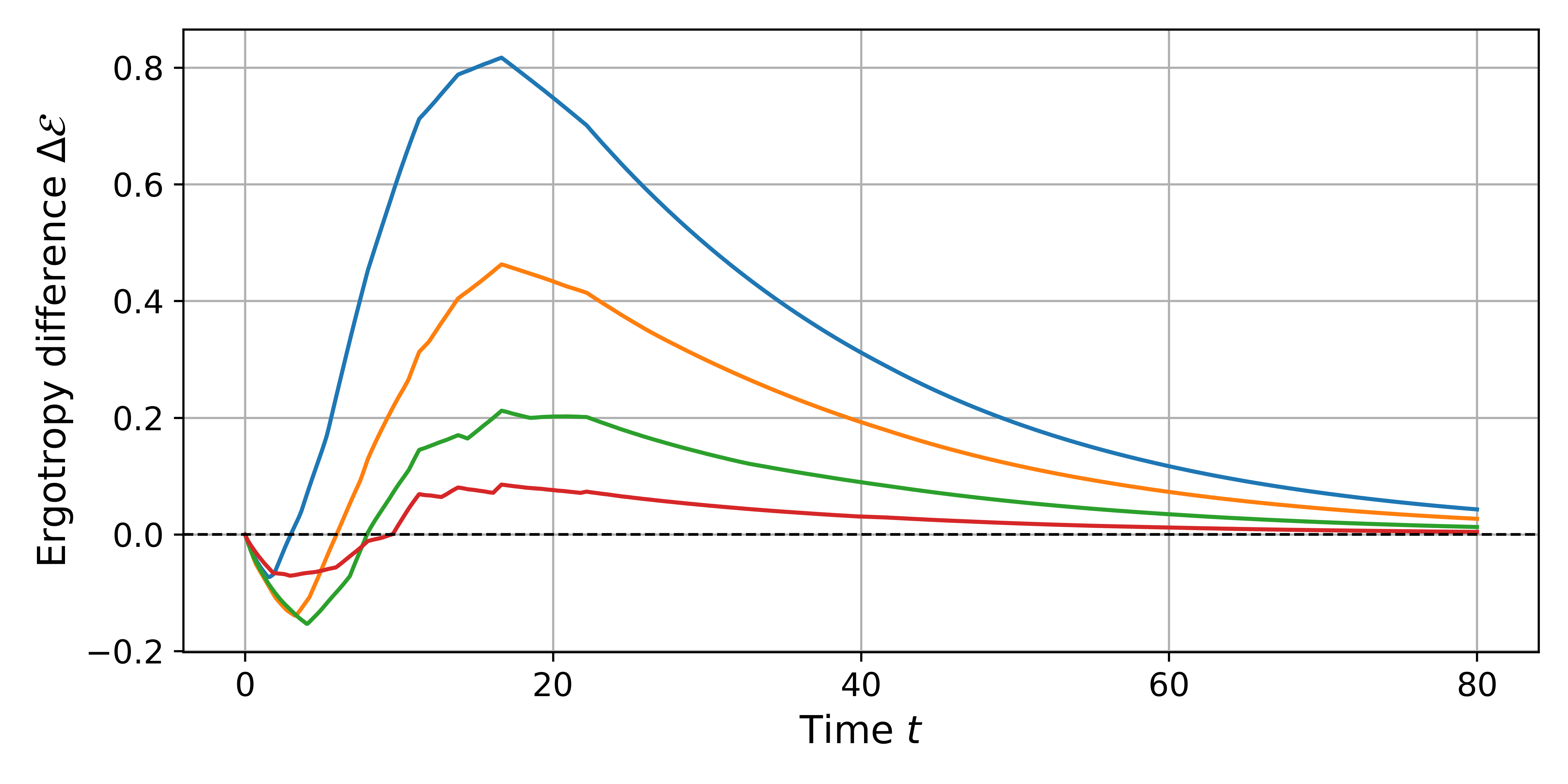}
    \caption{}
    \label{Four_Qubits_Ergotropies_Difference_Parallel}
  \end{subfigure}

  \caption{(a) Ergotropy $\mathcal{E}$ of the four-qubit system as function of time in presence of four parallel dissipative channels for $h = 0.1$, $\gamma = 0.05$ and initial state's temperature $\beta = 0.2$ (blue curve), $\beta = 0.5$ (orange curve), $\beta = 1$ (green curve), $\beta = 2$ (red curve) and $\beta = 5$ (purple curve). (b) Ergotropy difference $\Delta \mathcal{E}(t) = \mathcal{E}_\beta(t) - \mathcal{E}_{\beta = 5.0}(t)$ with respect to the reference case $\beta = 5$ as function of time for $\beta = 0.2$ (blue curve), $\beta = 0.5$ (orange curve), $\beta = 1.0$ (green curve) and $\beta = 2.0$ (red curve). Overall, local dissipation yields \textcolor{black}{what can be seen as an} ergotropic Mpemba-like effect, with hotter states temporarily overtaking colder ones before reaching a common steady state.}
  \label{fig:four_qubits_parallel}
\end{figure}

We now solve the dynamics for a chain of $N = 4$ sites, starting again from the parallel case $\alpha^{(-)}=0$. In panel (a) of Fig. \ref{fig:four_qubits_parallel} ergotropy curves for different values of $\beta$ are shown. \textcolor{black}{As predicted at the beginning of Sec. IIIA for a generic chain size $N$,} having multiple parallel individual dissipative channels results in obtaining the same non-passive steady state independently from the temperature of the initial thermal state. \textcolor{black}{In the case $N=4$, the} main difference with respect to the previous case lies in the transient regime, where various crossings between different curves appear. In particular, \textcolor{black}{in Fig.   \ref{fig:four_qubits_parallel}}, we observe that the ergotropy associated with the colder initial state ($\beta = 5$, purple curve) is progressively overtaken by the curves corresponding to higher temperatures, each crossing occurring at a different time in ergotropic Mpemba-like fashion~\cite{Medina25, Li25, teza2026speedups}. The crossings become even more evident in panel (b) of the same figure, where we plot the difference between the ergotropy at each temperature and that at $\beta = 5$, taken as a reference. In this representation, positive values indicate that the ergotropy for a given temperature is higher than in the reference case, whereas negative values indicate the opposite. The ergotropy crossings appear at those times such that $\Delta \mathcal{E}(t) \equiv \mathcal{E}_\beta(t) - \mathcal{E}_{\beta = 5.0}(t) = 0$. \textcolor{black}{This phenomenology is ultimately related to the fact that, due to the emergence of multiple crossings in
the energy spectrum, the passive energy becomes strongly non-monotonic in time. This is analogous to what discussed in Ref. \cite{Medina25} concerning the ergotropic Mpemba effect. More details supporting this analysis are reported in Appendix \ref{App:Four_Qubits_Crossings}.
Moreover, to further investigate the origin of these crossings, we have analyzed the decay modes of the Liouvillian governing the dynamics. Although the Liouvillian eigenvalues are independent of the initial temperature, the corresponding mode amplitudes depend on the initial state. We find that the slowest spectral modes have negligible overlap with the initial thermal states considered here, while the accessible decaying components exhibit essentially the same asymptotic decay rate for all temperatures. This indicates that the observed crossings, rather than originating from a conventional Mpemba mechanism involving different asymptotic relaxation rates or the suppression of slow Liouvillian modes, reflect the non-linear dependence of ergotropy on the state of the system, which allows the ordering of the extractable work to change even when the asymptotic relaxation rates are essentially identical. In this sense, the phenomenon is more appropriately characterized as an ergotropic Mpemba-like effect. Finally, we find that this effect persists for all the system sizes investigated, up to $N=9$. The numerical results are provided in Appendix \ref{App:Mpemba_39}.}

\begin{figure}[H]
  \centering

  \begin{subfigure}{\columnwidth}
    \centering
    \includegraphics[width=\columnwidth]{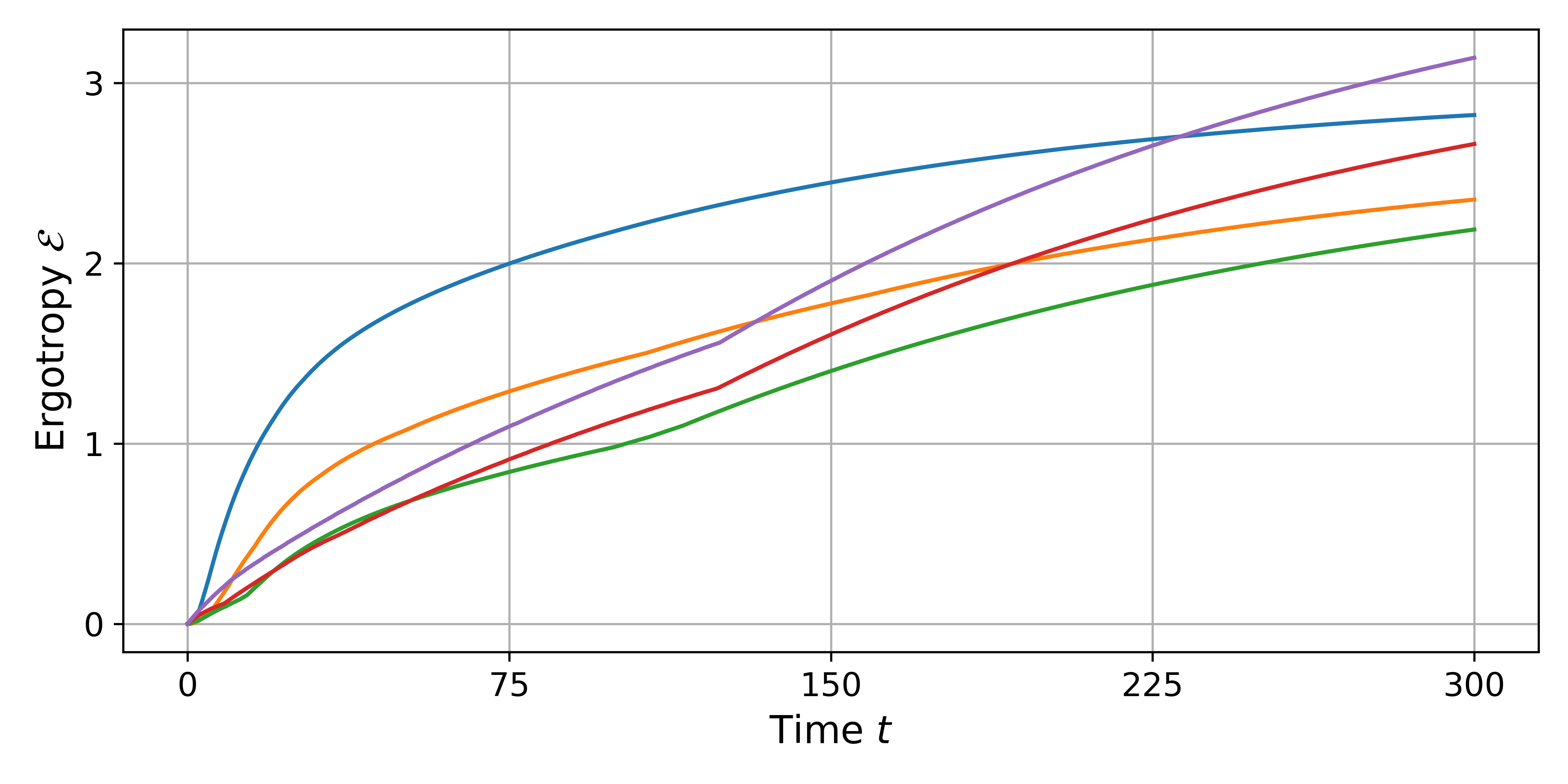}
    \caption{}
    \label{Four_Qubits_Ergotropies_Parallel}
  \end{subfigure}

  \vspace{0.4cm}

  \begin{subfigure}{\columnwidth}
    \centering
    \includegraphics[width=\columnwidth]{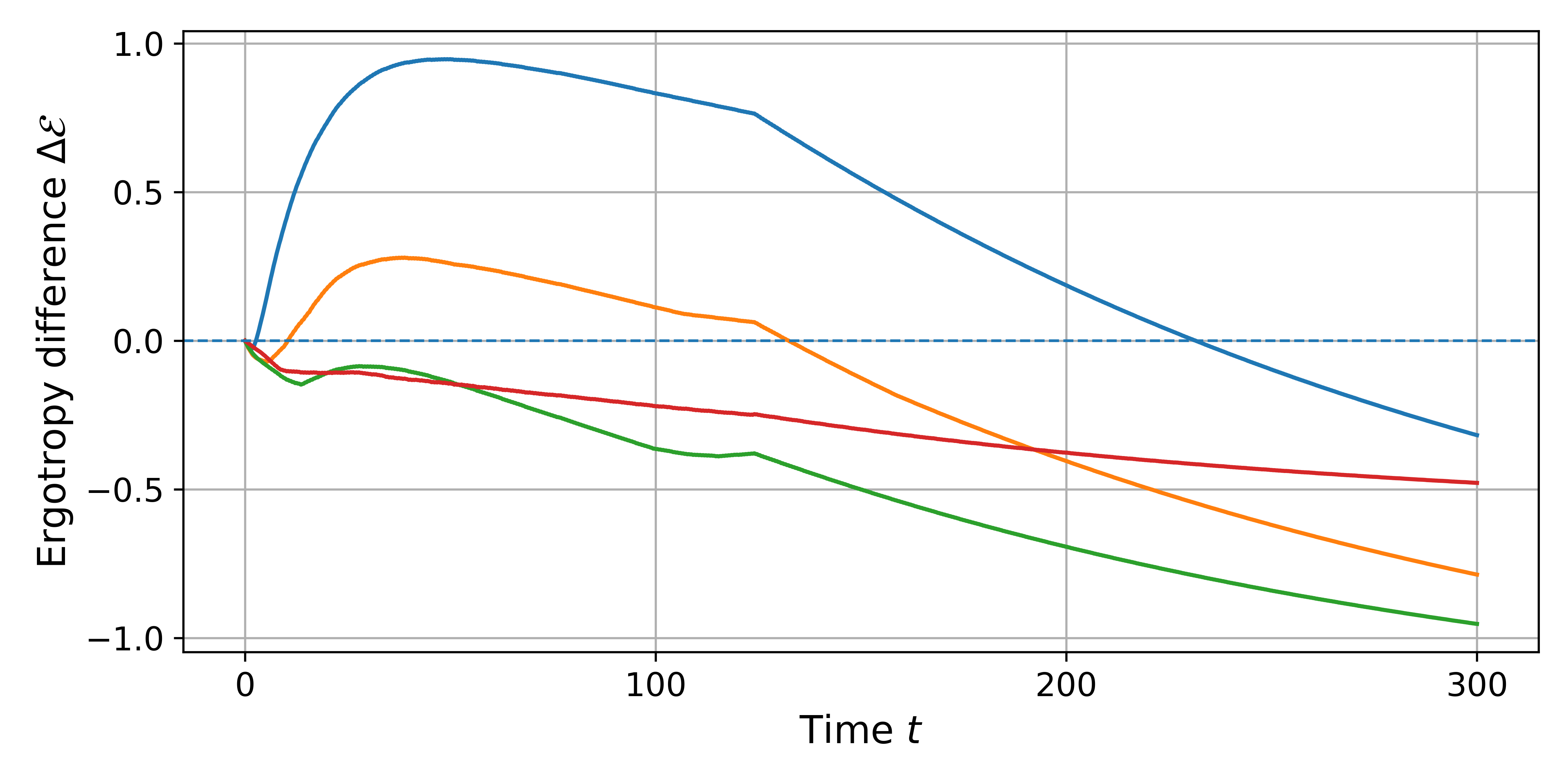}
    \caption{}
    \label{Four_Qubits_Ergotropies_Difference_Parallel}
  \end{subfigure}

  \caption{(a) Ergotropy $\mathcal{E}$ of the four-qubit system as function of time in presence of a collective dissipative channel for $h = 0.1$, $\gamma = 0.05$ and initial state's temperature $\beta = 0.2$ (blue curve), $\beta = 0.5$ (orange curve), $\beta = 1$ (green curve), $\beta = 2$ (red curve) and $\beta = 5$ (purple curve). (b) Ergotropy difference $\Delta \mathcal{E}(t) = \mathcal{E}_\beta(t) - \mathcal{E}_{\beta = 5.0}(t)$ with respect to the reference case $\beta = 5$ as function of time for $\beta = 0.2$ (blue curve), $\beta = 0.5$ (orange curve), $\beta = 1.0$ (green curve) and $\beta = 2.0$ (red curve). Collective dissipation confines the hot-state \textcolor{black}{ergotropic} advantage to a finite transient window and makes the stationary ergotropy dependent on the initial temperature.}
  \label{fig: four_qubits_collective}
\end{figure}

\noindent Now we \textcolor{black}{move to} the collective dissipation scenario. \textcolor{black}{In} both panels of Fig. \ref{fig: four_qubits_collective}, we observe that the ergotropy advantage of initially hotter states is restricted to a finite time window. Moreover, some ergotropy curves corresponding to different initial temperatures remain ordered throughout the entire evolution and do not cross at any time. Similarly to what we observed in the two-qubit system, considering a single collective dissipative bath results in having steady state values depending on the initial state's temperature. However, \textcolor{black}{differently from the two-qubit system where the steady state properties rely on the presence of a critical temperature $\beta_c$, the four-qubit steady state is non-passive for all the initial temperatures considered. Extending the analysis to additional chain lengths reveals a more complex scenario, in which the temperature dependence of the stationary ergotropy varies significantly with the system size. In particular, the trends observed for different values of $N$ do not appear to follow a simple universal law. To gain further insight into this behavior, we have performed additional numerical analyses. Since a steady state must satisfy $\mathcal{L}(\rho_{\mathrm{SS}})=0$, it is useful to identify a family of operators that are necessarily stationary. In particular, any dark energy eigenstate $|D_\alpha\rangle$, i.e. any state simultaneously satisfying
\begin{equation}
H|D_\alpha\rangle=E_\alpha|D_\alpha\rangle,
\qquad
S^-|D_\alpha\rangle=0,
\end{equation}
defines a stationary projector $|D_\alpha\rangle\langle D_\alpha|$ of the full Liouvillian. However, these projectors are not sufficient to describe the entire stationary operator space, so we investigate numerically whether additional stationary operators exist for the model considered here. Interestingly, for chain sizes $N=3,\ldots,9$, we identify a number of energy eigenstates satisfying the above conditions that follows the empirical relation
\begin{equation}
n_{\mathrm{dark}}(N)=
\left\lfloor\frac{N}{2}\right\rfloor+1,
\end{equation}
a number that coincides with the dimension of the kernel of the Liouvillian. This provides numerical evidence of the fact that, for the system sizes investigated, these projectors exhaust the stationary operator space. As a further consistency check, we evaluate the projection of the dynamically reached steady state onto the identified dark subspace. We find that
\begin{equation}
p_{\mathrm{dark}}^{\mathrm{SS}}
=\mathrm{Tr}[P_{\mathrm{dark}}\rho_{\mathrm{SS}}]
\simeq 1,
\end{equation}
with $P_{dark} = \sum_\alpha \ket{D_\alpha}\bra{D_\alpha}$, even though the corresponding initial projection $p_{dark}(0)$ is generally smaller than unity, showing that the dissipative dynamics drives the system towards the dark subspace. We can consequently write
\begin{equation}
\rho_{\mathrm{SS}}(\beta)
\simeq
\sum_{\alpha=1}^{n_{\mathrm{dark}}}
p_\alpha(\beta)
|D_\alpha\rangle\langle D_\alpha|,
\end{equation}
where the stationary weights $p_\alpha$ depend on the initial thermal distribution. We can now express its ergotropy directly in terms of the populations of the dark eigenstates. The average energy reads
\begin{equation}
E_{\mathrm{SS}}(\beta)
=
\sum_{\alpha=1}^{n_{\mathrm{dark}}}
p_\alpha(\beta)E_\alpha,
\end{equation}
whereas the passive energy is obtained by rearranging the stationary populations in decreasing order and assigning them to the lowest energy levels of the full Hamiltonian. Denoting by $p_{(k)}(\beta)$ the $k$-th largest stationary population and by $E_k^\uparrow$ the $k$-th lowest eigenvalue of $H$, we obtain
\begin{equation}
E_{\mathrm{pas,SS}}(\beta)
=
\sum_{k=1}^{n_{\mathrm{dark}}}
p_{(k)}(\beta)E_k^\uparrow,
\end{equation}
so that the stationary ergotropy can be written as
\begin{equation}
\mathcal{E}_{\mathrm{SS}}(\beta)
=
\sum_{\alpha=1}^{n_{\mathrm{dark}}}
p_\alpha(\beta)E_\alpha
-
\sum_{k=1}^{n_{\mathrm{dark}}}
p_{(k)}(\beta)E_k^\uparrow.
\end{equation}
For all the initial temperatures investigated and for $N=3,\ldots,9$, the stationary states are found to be non-passive. However, their ergotropic behavior exhibits a strong dependence on the system size. For instance, for $N=3$ the stationary ergotropy is independent of the initial temperature, despite the fact that the average and passive energies separately vary with $\beta$. For $N=4$, instead, the ergotropy displays a non-monotonic dependence, reaching a minimum at intermediate temperatures before increasing towards the low-temperature regime. Different behaviors emerge for larger chains, where the stationary ergotropy can also increase or decrease monotonically, indicating that no simple universal trend with respect to the initial temperature can be identified from the system sizes investigated. A closer inspection of the stationary populations of the individual dark eigenstates $\ket{D_\alpha}$ reveals that, for a fixed $N$, increasing the initial inverse temperature generally leads to a distribution increasingly biased towards one of the dark eigenstates, although the degree of concentration depends on the system size itself. However, identifying which specific dark eigenstate acquires the largest weight appears to be nontrivial and cannot be simply inferred from its energy.} 

\section{RESULTS ON DEPHASING CHANNEL} \label{Sec:alpha_one}

\begin{figure}[H]
  \centering

  \begin{subfigure}{\columnwidth}
    \centering
    \includegraphics[width=\columnwidth]{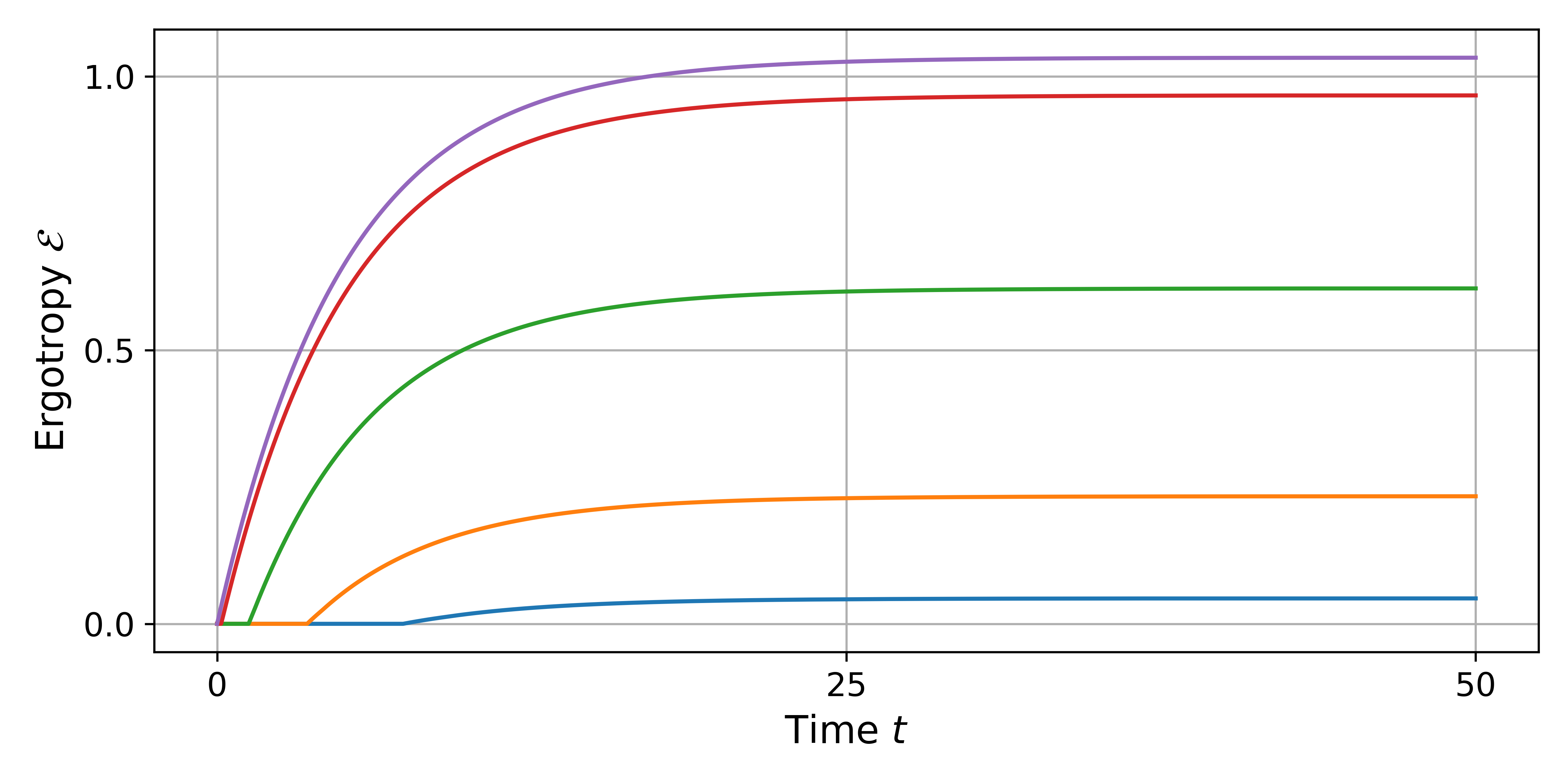}
    \caption{}
  \end{subfigure}

  \vspace{0.4cm}

  \begin{subfigure}{\columnwidth}
    \centering
    \includegraphics[width=\columnwidth]{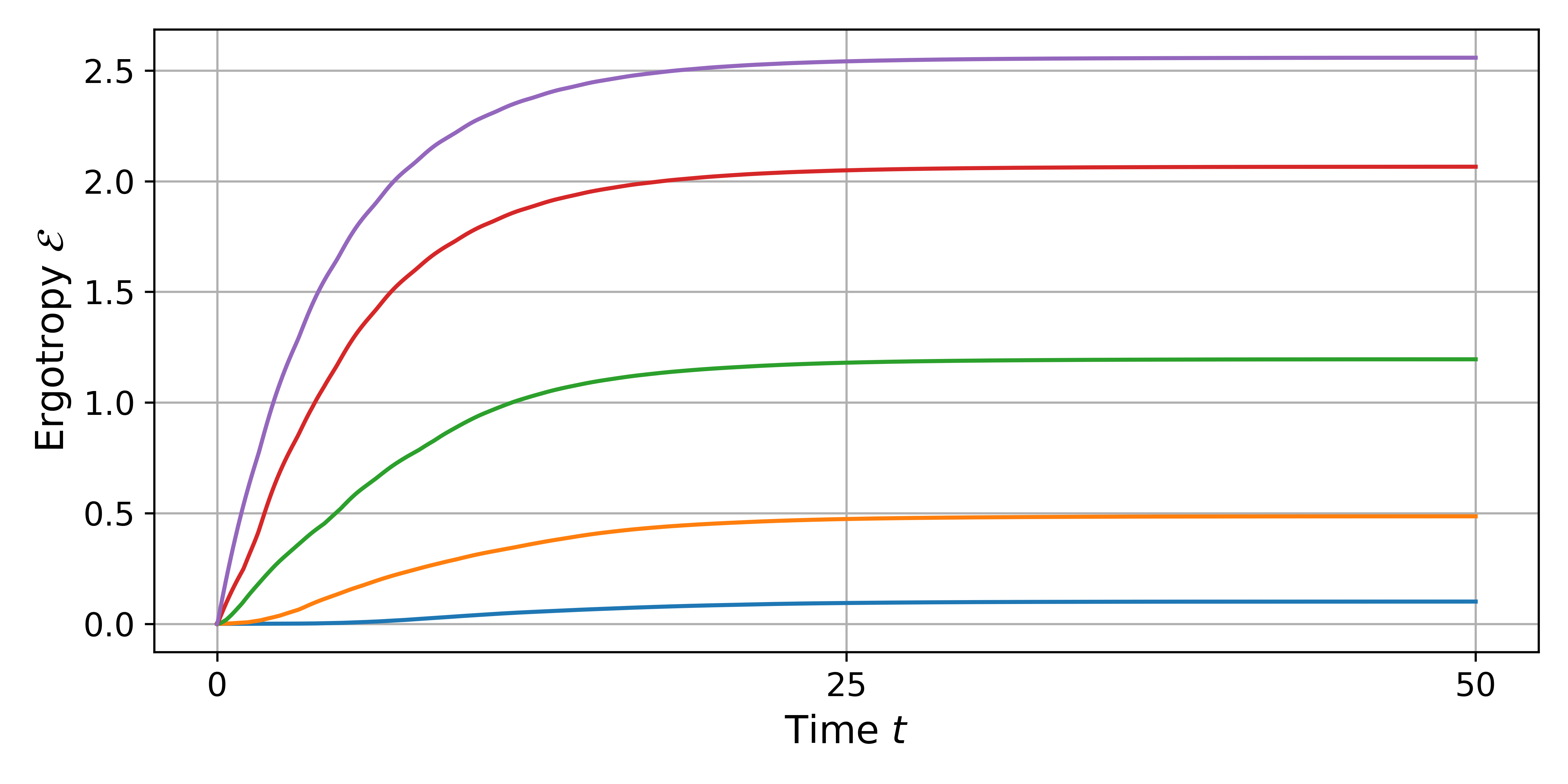}
    \caption{}
  \end{subfigure}

  \caption{Ergotropy $\mathcal{E}$ of the two-qubit (a) and the four-qubit (b) system as function of time in presence of parallel dephasing channels for $h = 0.1$, $\gamma = 0.05$ and initial state's temperature $\beta = 0.2$ (blue curve), $\beta = 0.5$ (orange curve), $\beta = 1$ (green curve), $\beta = 2$ (red curve) and $\beta = 5$ (purple curve). \textcolor{black}{We observe that the coldest initial state is characterized by the largest ergotropy at all times, with no crossing among the curves.}}
  \label{fig:dephasing}
\end{figure}

We \textcolor{black}{consider now} the full dephasing scenario, obtained by fixing $\alpha = 1$ in Eq. \eqref{Full_Superoperator}. Starting from the parallel case $\alpha^{(z)} = 0$ and addressing first the two-qubit system, we observe that a generic element of the superoperator $\mathcal{D}^{(z)}[\rho]$ can be written in the form
\textcolor{black}{\begin{equation} \label{Elementwise_dephasing}
    \big(\mathcal{D}^{(z)}[\rho]\big)_{a b}=\gamma\sum_{i=1}^2\big[z_i(a)z_i(b)-1\big]\rho_{ab},
\end{equation}
where $z_i(a) \in \{-1,1\}$ is the eigenvalue of $\sigma_i^z$ on the basis vector $\ket{a}$. Eq. \eqref{Elementwise_dephasing} shows that (i) for diagonal elements $a=b$ the dissipator vanishes, so that pure dephasing does not directly affect the populations, and (ii) for an off-diagonal element $\rho_{ab}$ each qubit, on which $\ket{a}$ and $\ket{b}$ differ, contributes with} a factor $-2\gamma$, so the total decay rate for that coherence is $-2\gamma n$ where $n$ is the number of differing qubits (e.g.\ single-qubit coherences decay as $\exp(-2\gamma t)$, while coherences differing on both qubits decay as $\exp(-4\gamma t)$). Using the same notation as in the previous Section, we obtain the following closed system of differential equations
\begin{equation}
\begin{cases}
& \dot p_{gg} = 0,\\
& \dot p_{eg} = -2i\big(c^*-c\big),\\
& \dot p_{ge} = -2i\big(c-c^*\big),\\
& \dot p_{ee} = 0,\\
& \dot c = -2i\big(p_{ge}-p_{eg}\big)-4\gamma\,c.
\end{cases}
\end{equation}
\textcolor{black}{An important consequence of this dynamics is that local dephasing preserves the total number of excitations, since both the XX Hamiltonian and the local dephasing operators commute with it. So, while the Hamiltonian can redistribute excitations among different sites through coherent hopping, local dephasing progressively suppresses the coherences underlying these processes. As a result, the total population associated with each excitation sector remains constant, even if the distribution among individual configurations in a given sector may change. Since these conserved sector populations depend on the initial temperature, the dynamics retains a memory of the initial thermal state, as can be seen in} the plot shown in Fig. \ref{fig:dephasing} for a two-qubit, panel (a), and a four-qubit system, panel (b). In this scenario, starting from the coldest state always results in having the highest amount of ergotropy and no Mpemba-like crossings are present.

\noindent We conclude this part considering the $\alpha^{(z)} = 1$ case, corresponding to a single collective dephasing channel. The dephasing superoperator takes the form
\begin{equation}
    \mathcal{D}^{(z)}[\rho] = \gamma \left(S_z \rho S_z-\frac{1}{2}\left\{S_z^2,\rho\right\}\right).
\end{equation}
with $S_z = \sum_{i=1}^N \sigma_i^z$. \textcolor{black}{Unlike local dephasing, the collective channel couples to the total magnetization $S_z$ and therefore suppresses only coherences between states belonging to different magnetization sectors, leaving those within the same sector unaffected. For instance, the coherence between $|eg\rangle$ and $|ge\rangle$ is destroyed by local dephasing, whereas it is preserved by collective dephasing, since these two states have the same total magnetization. In our case, the initial Gibbs state satisfies $[\rho_\beta,S_z]=0$, because it is a function of the Hamiltonian and $[H,S_z]=0$. Consequently, the collective dephasing dissipator vanishes on $\rho_\beta$. Moreover, since $[H,\rho_\beta]=0$, the initial thermal state is also stationary under the full Liouvillian dynamics. Therefore, the system remains in its initial Gibbs state throughout the evolution. Since Gibbs states are completely passive with respect to their Hamiltonian, the ergotropy remains identically zero at all times~\cite{Pusz78}.}
\section{Conclusion}

In this paper, we investigated work extraction in open quantum batteries composed of interacting spin chains weakly coupled to engineered environments. Focusing \textcolor{black}{mainly} on \textcolor{black}{the analytically treatable cases of} two- and four-qubit XX systems initially prepared in thermal Gibbs states, we analyzed how dissipation and dephasing, acting either locally or collectively, affect the evolution of ergotropy during both transient and steady-state dynamics.

We showed that purely dissipative dynamics can activate finite ergotropy starting from completely passive thermal states. In the presence of parallel dissipation channels, ergotropy is generated transiently and the system eventually reaches a non-passive steady state that is independent of the initial temperature \textcolor{black}{and whose stationary ergotropy scales linearly with $N$}. In this regime, for the four-qubit chain, we identified temperature-dependent crossings of ergotropy curves, \textcolor{black}{similar to the ergotropic Mpemba effect,} in which initially hotter states can temporarily outperform colder ones in terms of extractable work. \textcolor{black}{We also find out numerically that these crossings appear at least until nine-qubit long chains.}

In contrast, collective dissipation leads to qualitatively different behavior. For two qubits, the passivity of the steady state depends on the initial temperature, with a critical inverse temperature separating passive and non-passive regimes. \textcolor{black}{For $N$ qubits, with $N \in [3,9]$, we found that the stationary states are entirely supported, within numerical precision, on a subspace spanned by energy eigenstates annihilated by the collective lowering operator. Moreover, the stationary populations of these states depend non-trivially on both the initial temperature and the system size.} 

Finally, we demonstrated that \textcolor{black}{environments leading to dephasing are fundamentally different} from dissipative ones. Collective dephasing suppresses ergotropy generation, while the parallel one does not present transient advantages for the hotter initial states. This highlights the essential role of population reshuffling induced by dissipation, as opposed to mere \textcolor{black}{dephasing}, in enabling work extraction assisted by the environment.

Overall, our results clarify how the properties of the environment and the nature of the noise channels shape the ergotropic properties of open quantum batteries, \textcolor{black}{at least for} a small number of qubits. These findings provide \textcolor{black}{an important} step toward understanding how environmental engineering can be exploited as a resource for quantum battery charging in purely dissipative protocols. \textcolor{black}{Several interesting directions for future investigations emerge from our results. A first natural direction concerns the collective-dissipation scenario, whose stationary properties appear to exhibit a rich dependence on both the system size and the initial temperature, and therefore deserve further study. Then, a systematic understanding of the interplay between the spectral properties of the system Hamiltonian and the structure of the dissipative channels could provide a more general framework for characterizing dissipative ergotropy activation. In this work, we have explicitly addressed the role of energy-level crossings in the two-qubit model, showing how the change in the ordering of the singlet and fully polarized states determines whether local dissipation can lead to a non-passive steady state. A natural extension would be to investigate systems with more complex energy spectra, where multiple level crossings and changes in the ground-state structure may give rise to different regimes of ergotropy activation and distinct transitions between passive and non-passive steady states. Other possible directions include investigating how these effects depend on the structure of the jump operators, as well as extending the analysis to the strong-coupling regime, where non-Markovian effects may become relevant.}

\begin{acknowledgements}
D. Farina acknowledges
financial support from PNRR MUR Project No.
PE0000023-NQSTI.
\end{acknowledgements}

\appendix
\section{Link with the Jaynes-Cummings model} \label{App:JC}

Here, we show that our engineered dissipative jump operators can be obtained starting from a Jaynes-Cummings (JC) model in the limit of lossy cavity. The JC model describes the interaction between a two-level atom and a single quantized mode of an optical cavity~\cite{Schleich2015}. Within the rotating-wave approximation, \textcolor{black}{this} system is described by the Hamiltonian
\begin{equation}
    H_{JC} = \frac{\omega_q}{2} \sigma_z + \omega_c a^\dagger a 
    + g(\sigma_- a^\dagger + \sigma_+ a),
\end{equation}
where $\omega_q$ and $\omega_c$ denote the qubit and cavity frequencies, respectively, $g$ is the light-matter coupling strength, and $a$ ($a^\dagger$) is the bosonic annihilation (creation) operator. Cavity losses related to the JC model can be described by a phenomenological master equation for the generic state of qubit+cavity radiation $\rho_{JC}$ of the form \cite{Scala2007}
\begin{equation}
    \dot{\rho}_{JC} = -i[H_{JC}, \rho_{JC}] + k\left(a \rho_{JC} a^\dagger - \frac{1}{2} \{a^\dagger a, \rho_{JC}\}\right)
\end{equation}
with $k$ representing the rate of loss of photons from the cavity. The lossy cavity limit we are interested in corresponds to the condition
\begin{equation} \label{Limit_math}
    \frac{k}{g}>>1.
\end{equation}
This limit states that photons emitted by the atom leave the cavity almost istantaneously before interacting with anything else. Denoting with $\ket{\delta,n}$ the tensor product between the atom state $\ket{\delta}$ (with $\delta = g,e$ if the atom is respectively in the ground or excited state) and the cavity state $\ket{n}$ (with $n$ indicating the number of photons in the cavity), in this limit only the $n=0$ and $n=1$ sectors are relevant, i.e. a very lossy cavity doesn't contain two or more photons at the same time. According to this, the Hamiltonian of the JC model can be written as a 4x4 matrix in the basis $\{\ket{g,0}, \ket{e,0}, \ket{g,1}, \ket{e,1}\}$
\begin{equation}
    H_{JC} = \begin{pmatrix}
        -\frac{\omega_q}{2} & 0 & 0 & 0 \\
        0 & \frac{\omega_q}{2} & g & 0 \\
        0 & g & \omega_c - \frac{\omega_q}{2} & 0 \\
        0 & 0 & 0 & \omega_c + \frac{\omega_q}{2}
    \end{pmatrix}.
\end{equation}
In the large-loss limit of Eq.~\eqref{Limit_math}, the cavity can be adiabatically eliminated. Physically, this regime corresponds to a separation of timescales in which the cavity relaxes much faster than the typical time associated to the atom--cavity interaction. As a consequence, the cavity mode follows the atomic dynamics quasi-instantaneously and can be treated as a fast degree of freedom slaved to the atom. Eliminating this fast variable leads to an effective reduced description involving only the atomic degrees of freedom, yielding the following 2x2 master equation for the atom:
\begin{equation}
    \dot{\rho}_a = -i [H_\mathrm{LS}, \rho_a] 
    + \Gamma_\mathrm{eff} \left(\sigma_- \rho_a \sigma_+ - \frac{1}{2}\{\sigma_+ \sigma_-, \rho_a\}\right),
    \label{reduced_atom}
\end{equation}
where $\sigma_-$ is the atomic lowering operator, 
\begin{equation}
    \Gamma_\mathrm{eff} = \frac{4 g^2 \kappa}{\kappa^2 + 4 \Delta^2}
\end{equation}
is the effective decay rate, and 
\begin{equation}
    H_\mathrm{LS} = -\frac{g^2 \Delta}{\frac{k^2}{4} + \Delta^2} |e\rangle\langle e|
\end{equation}
accounts for the Lamb shift ($\Delta = \omega_c - \omega_q$). Importantly, the derivation of the effective master equation is independent of the initial state of the atom. Any atomic state, whether pure, mixed, or thermal, can be used as the initial condition, and the adiabatic elimination procedure still yields the same form of the effective dynamics. The specific choice of the initial state only affects the resulting population dynamics $\rho_{gg}(t)$ and $\rho_{ee}(t)$, but does not alter the structure of the master equation or the effective decay rate $\Gamma_\mathrm{eff}$.

\begin{figure}[H]
    \centering
    \includegraphics[width=\columnwidth]{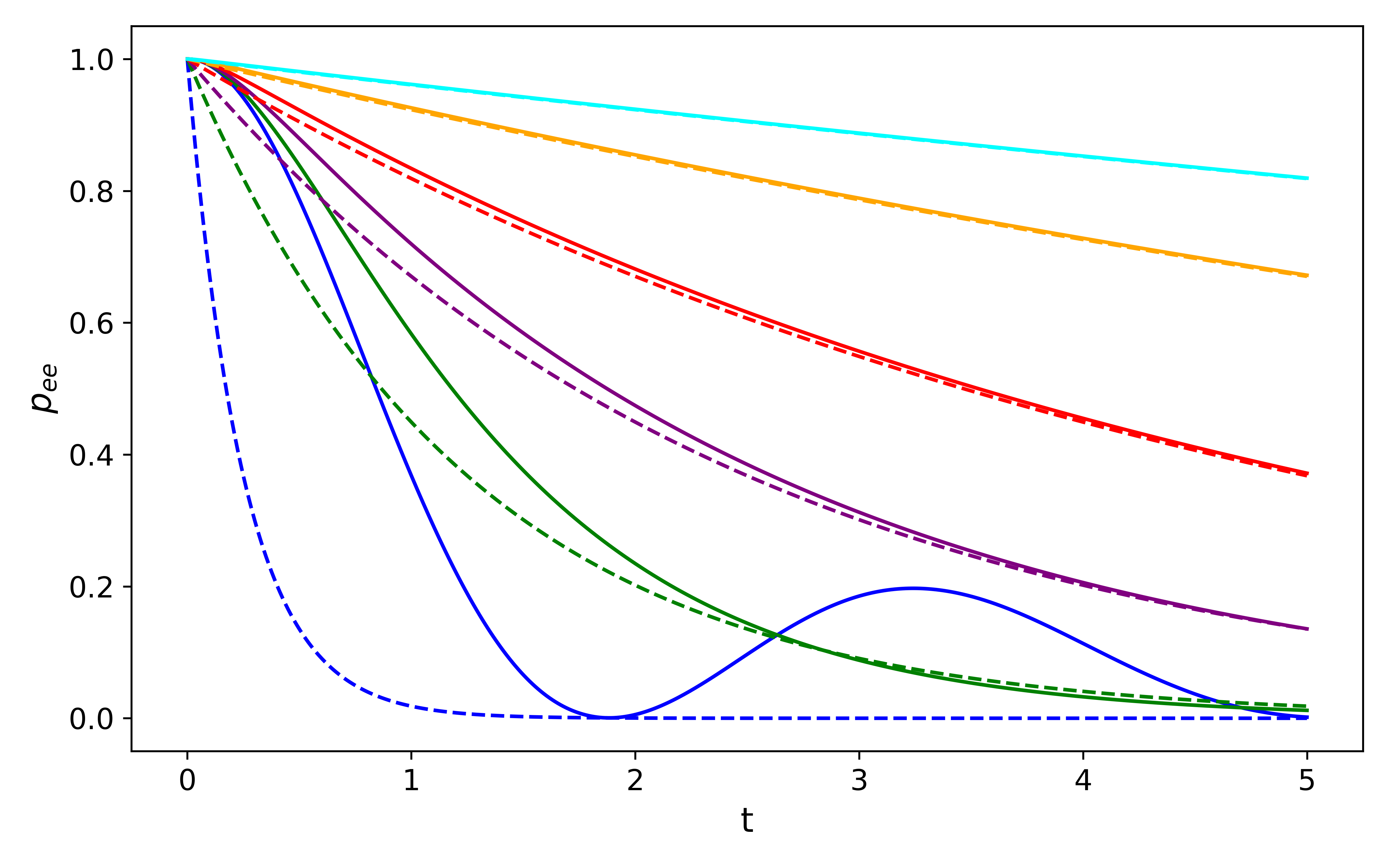}
    \caption{Comparison between the full 4x4 Jaynes-Cummings evolution with cavity dissipation (solid lines) and the effective 2x2 atomic master equation (dashed lines) for $k/g = 1$ (blue curve), $k/g = 5$ (green curve), $k/g = 10$ (purple curve), $k/g = 20$ (red curve), $k/g = 50$ (orange curve) and $k/g = 100$ (cyan curve). The effective Lindblad description reproduces the full lossy-cavity dynamics in the large-loss limit.}
    \label{fig_app:jc_vs_effective}
\end{figure}

To verify the validity of this effective model, we perform a direct comparison between the full 4x4 JC evolution with cavity dissipation and the reduced 2x2 atomic master equation in Eq. (\ref{reduced_atom}). In Fig. \ref{fig_app:jc_vs_effective}, we show as an example, the time evolution of each matrix element for several values of the cavity loss rate $\kappa$ for an atom-cavity system in the initial state $\ket{e,0}$. The solid lines correspond to the full 4x4 simulation followed by a partial trace over the cavity, while the dashed lines correspond to the evolution under the effective 2x2 master equation. For the initial state considered here the atomic density matrix remains diagonal during the evolution, i.e. $p_{eg}=p_{ge}=0$ at all times. Physically, this follows from the fact that the Jaynes-Cummings interaction and the dissipative cavity channel do not generate coherences between the atomic energy eigenstates when starting from a diagonal state. As a consequence, the atomic dynamics is fully characterized by the populations $p_{ee}$ and $p_{gg}$, which satisfy $p_{ee}+p_{gg}=1$. For this reason, in the figure we only display the evolution of $p_{ee}$. As expected, the agreement between the two approaches improves as $\kappa/g$ increases, providing a numerical demonstration that the engineered dissipative dynamics emerges naturally from the microscopic JC model in the large-loss limit. Finally, we briefly mention that the analogous comparison for the collective dissipative channel requires having all qubits inside the same cavity, extending the JC model to the Tavis-Cummings one~\cite{Schleich2015}.

\section{Interpolating cases} \label{App:Interpol}
In this Appendix we report the results arising from the interpolation of the various \textcolor{black}{regimes discussed in the main text}.
\subsection{Dissipative case}
\begin{figure}[H]
  \centering

  \begin{subfigure}{\columnwidth}
    \centering
    \includegraphics[width=0.8\columnwidth]{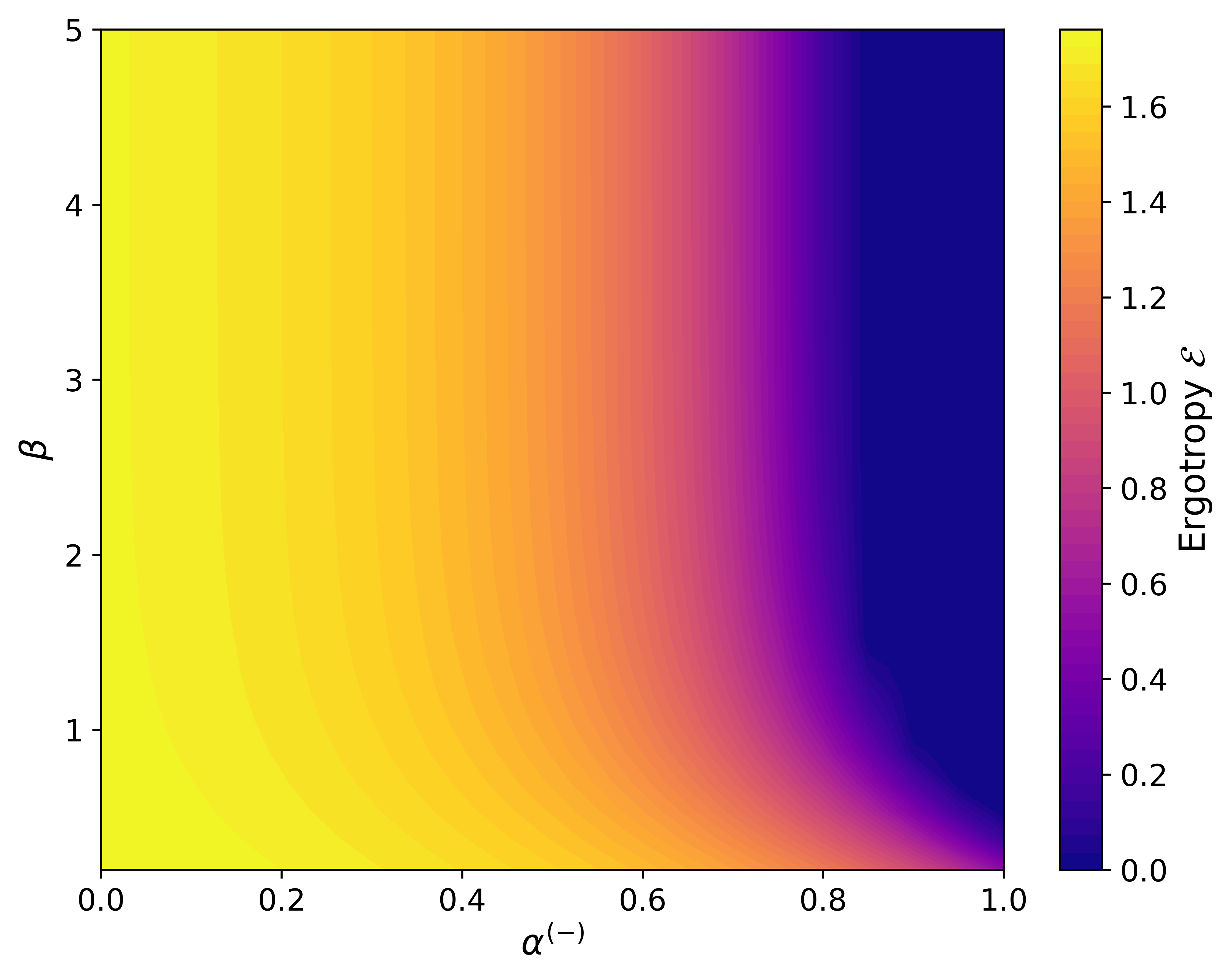}
    \caption{}
  \end{subfigure}

  \vspace{0.4cm}

  \begin{subfigure}{\columnwidth}
    \centering
    \includegraphics[width=0.8\columnwidth]{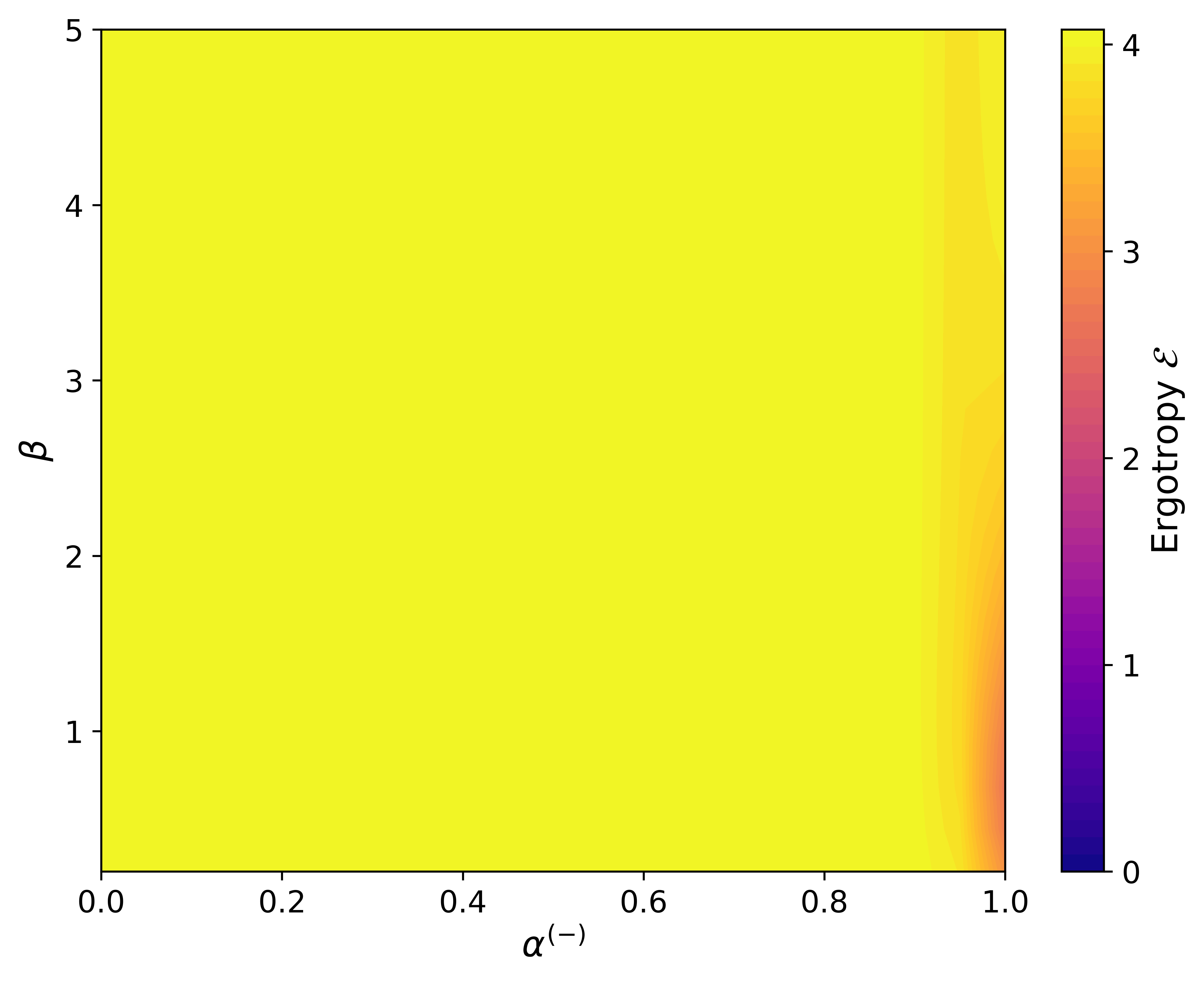}
    \caption{}
  \end{subfigure}

  \caption{Ergotropy $\mathcal{E}$ of the system's steady state as function of $\alpha^{(-)}$ and $\beta$ for $h = 0.1$, $\gamma = 0.05$ and $t = 800$ for (a) $N = 2$ qubits and (b) $N = 4$ qubits. Overall, increasing collectivity drives the two-qubit dissipative steady state toward passivity, while the four-qubit case remains largely non-passive.}
  \label{fig_app:interpolation_diss}
\end{figure}

In Fig. \ref{fig_app:interpolation_diss} we fix $\alpha = 0$ in Eq. \eqref{Full_Superoperator} and we plot the ergotropy of the steady state varying $\alpha^{(-)}$, which represents the degree of collectivity of the channels, and the inverse temperature $\beta$. For $\alpha^{(-)} = 0$ and $\alpha^{(-)} = 1$ we recover the results already discussed in the main text. For intermediate values of $\alpha^{(-)}$ we observe two different \textcolor{black}{behaviors} for the two-qubit, panel (a), and the four-qubit chain, panel (b). In the former case, increasing the collectivity of the channels drives the steady state towards passivity, and the colder the initial state, the
earlier this passivity is reached. Interestingly, a fully collective dissipation is not required for this behavior to emerge, since this trend already appears for values $\alpha^{(-)} \gtrsim 0.8$. In contrast, in the latter case a passive steady state is never reached indicating that the four-qubit chain is much more resilient to bath collectivity, since up to at least \textcolor{black}{$\alpha^{(-)}  \approx 0.9$} the steady-state ergotropy always coincides with that of the local case.

\subsection{Dephasing case}
\begin{figure}[H]
  \centering

  \begin{subfigure}{\columnwidth}
    \centering
    \includegraphics[width=0.8\columnwidth]{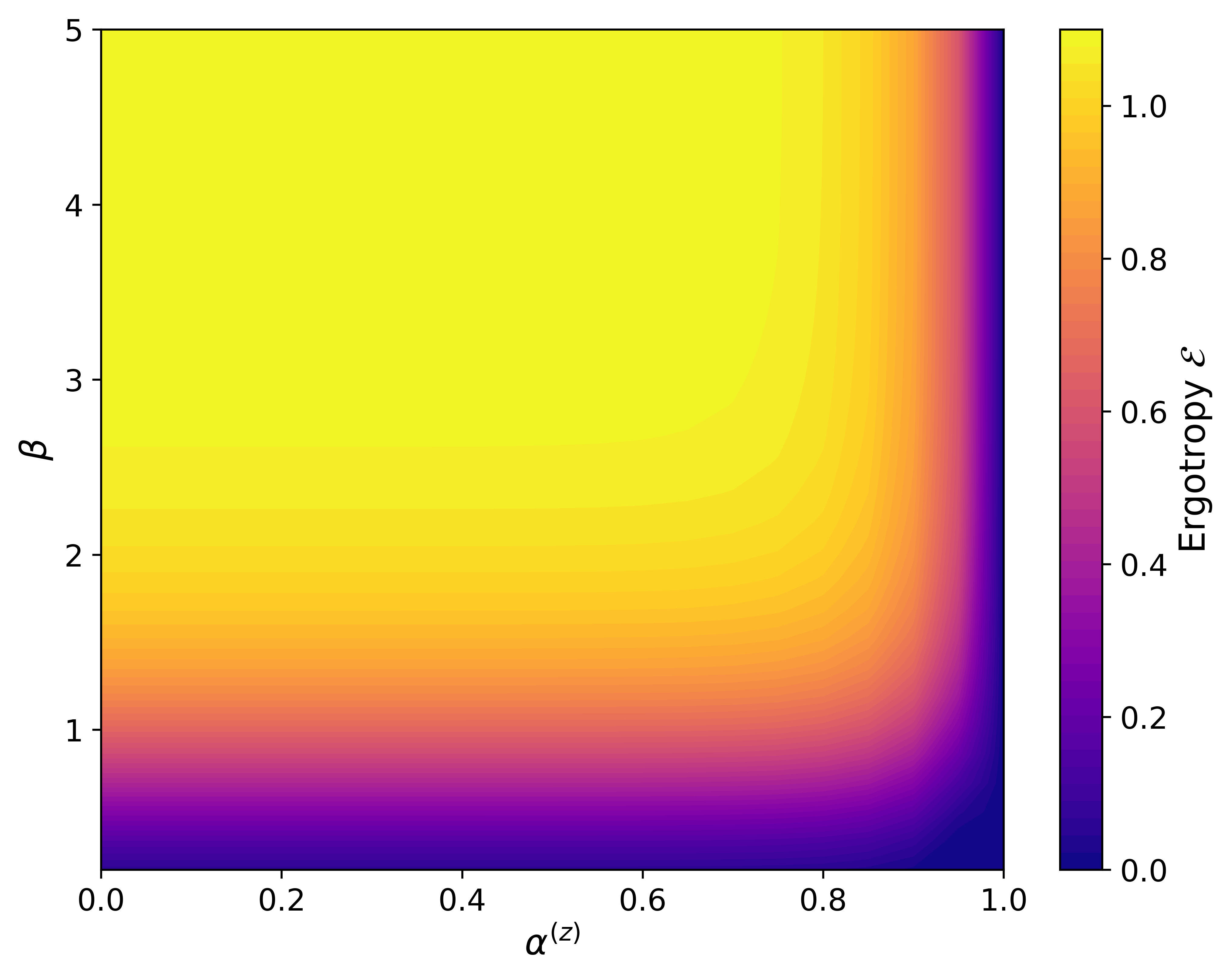}
    \caption{}
  \end{subfigure}

  \vspace{0.4cm}

  \begin{subfigure}{\columnwidth}
    \centering
    \includegraphics[width=0.8\columnwidth]{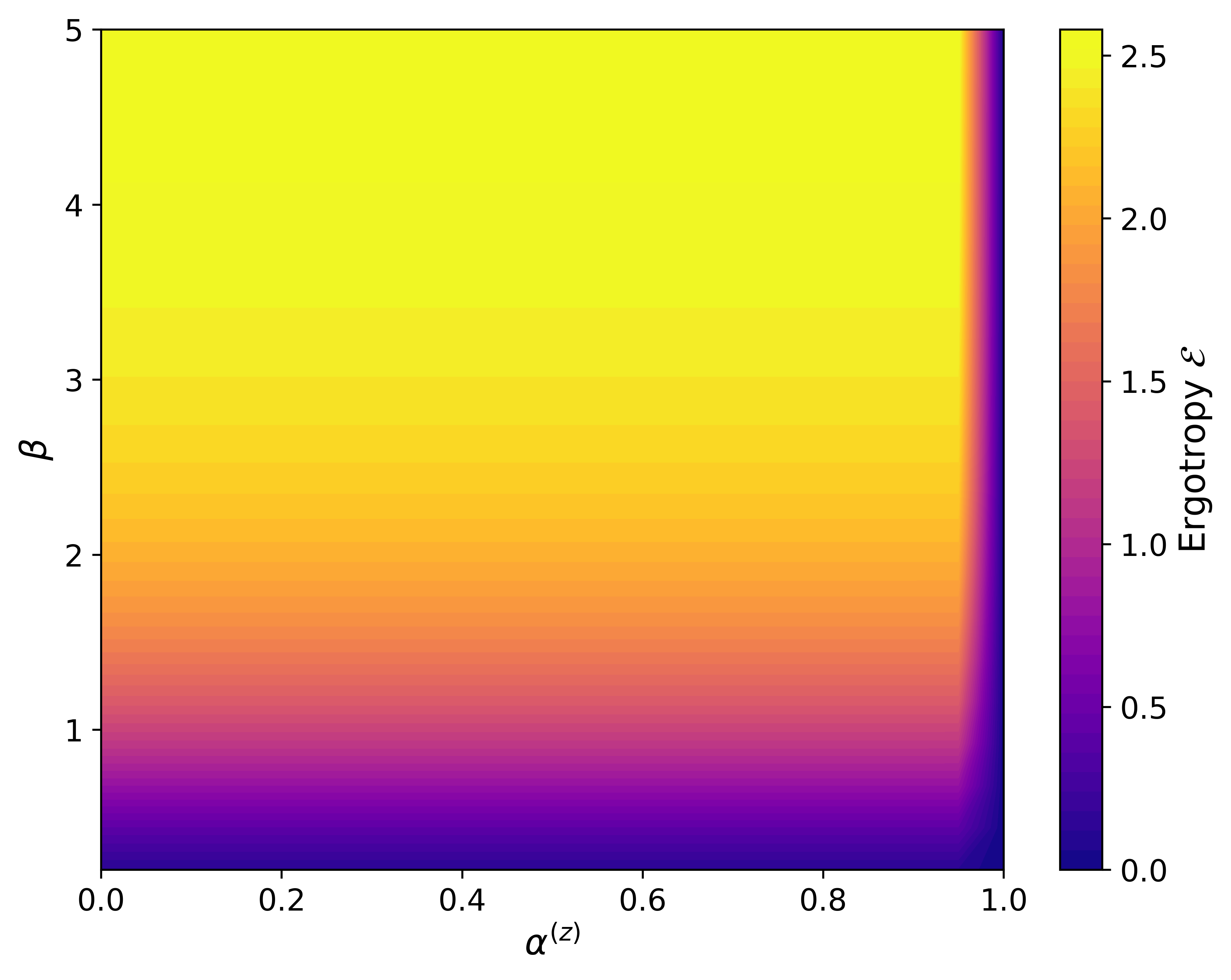}
    \caption{}
  \end{subfigure}

  \caption{Ergotropy $\mathcal{E}$ of the system's steady state as function of $\alpha^{(z)}$ and $\beta$ for $h = 0.1$, $\gamma = 0.05$ and $t = 800$ for (a) $N = 2$ qubits and (b) $N = 4$ qubits. In this scenario, changing the collectivity of the dephasing channel has only a minor effect on the steady-state ergotropy until $\alpha^{(z)} \simeq 1$.}
  \label{fig_app:interpolation_deph}
\end{figure}

We now fix $\alpha = 1$ in Eq. \eqref{Full_Superoperator} and we plot in Fig. \ref{fig_app:interpolation_deph} the ergotropy of the steady state varying $\alpha^{(z)}$ and $\beta$. Confirming what already observed in Sec. \ref{Sec:alpha_one}, for the full dephasing scenario the differences between the two-qubit, panel (a), and four-qubit chain, panel (b), are less pronounced.


\subsection{Channels interpolation}
\begin{figure*}[t]
    \centering
    \includegraphics[width=0.32\textwidth]{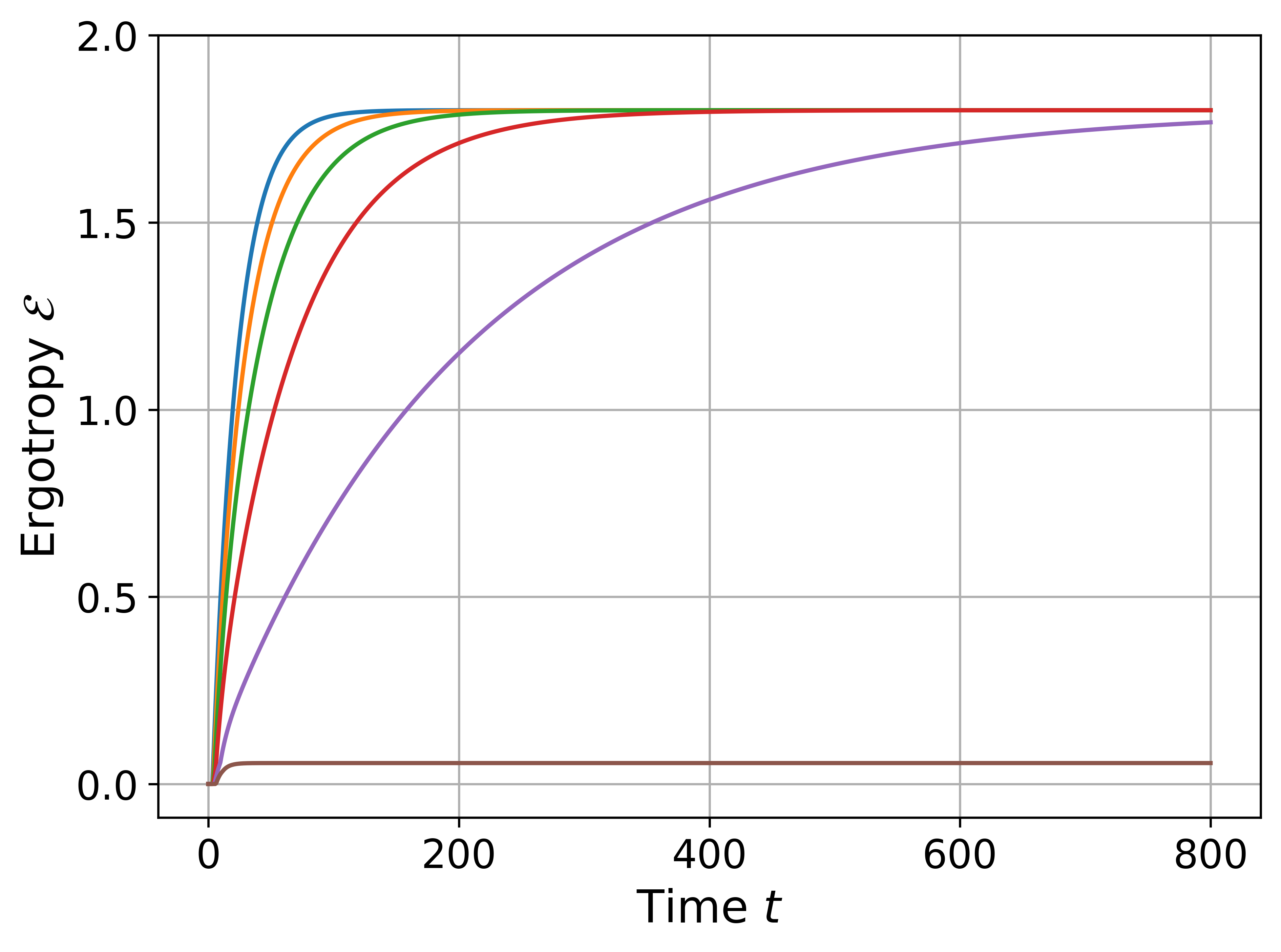}
    \includegraphics[width=0.32\textwidth]{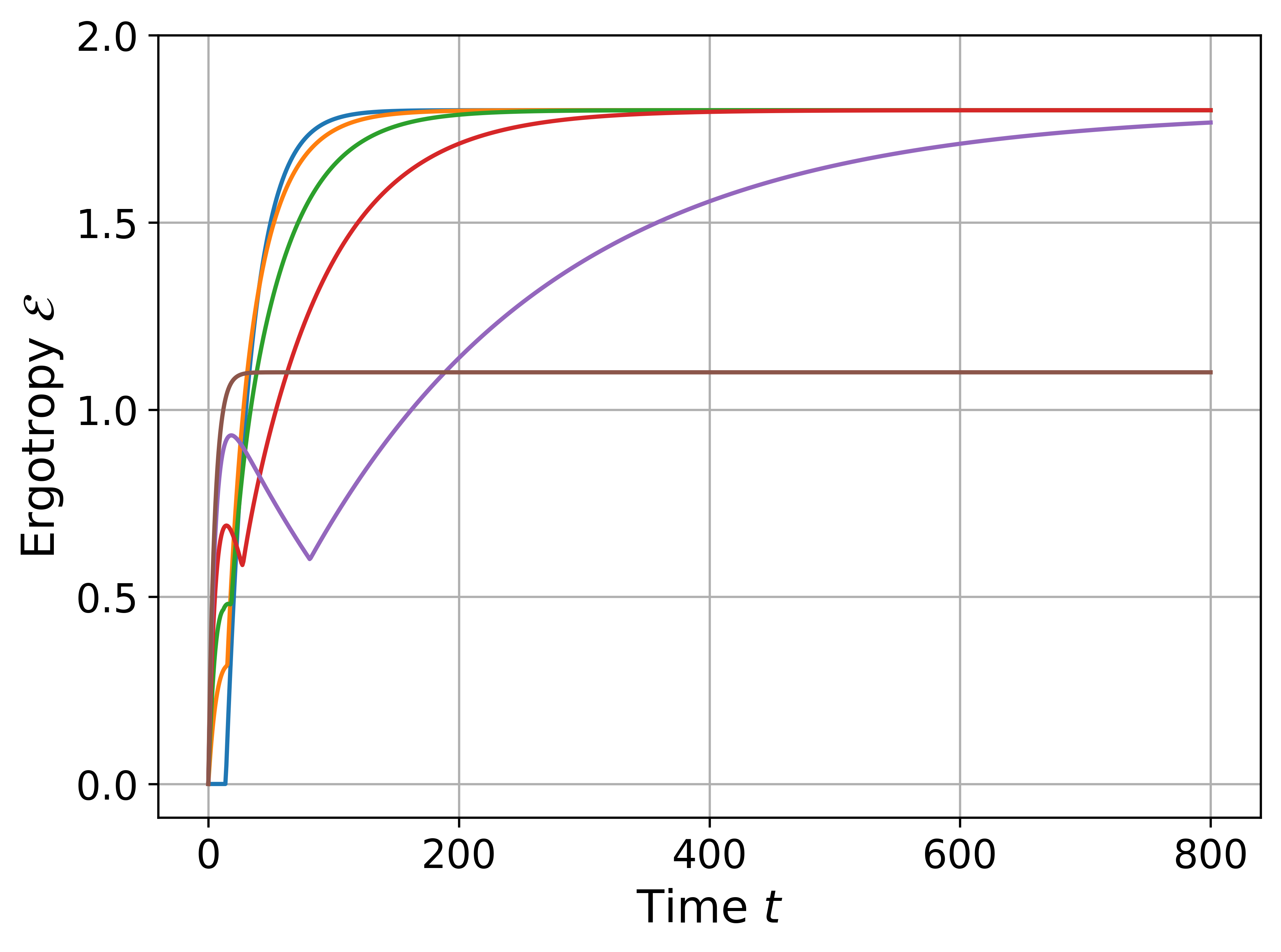}
    \includegraphics[width=0.32\textwidth]{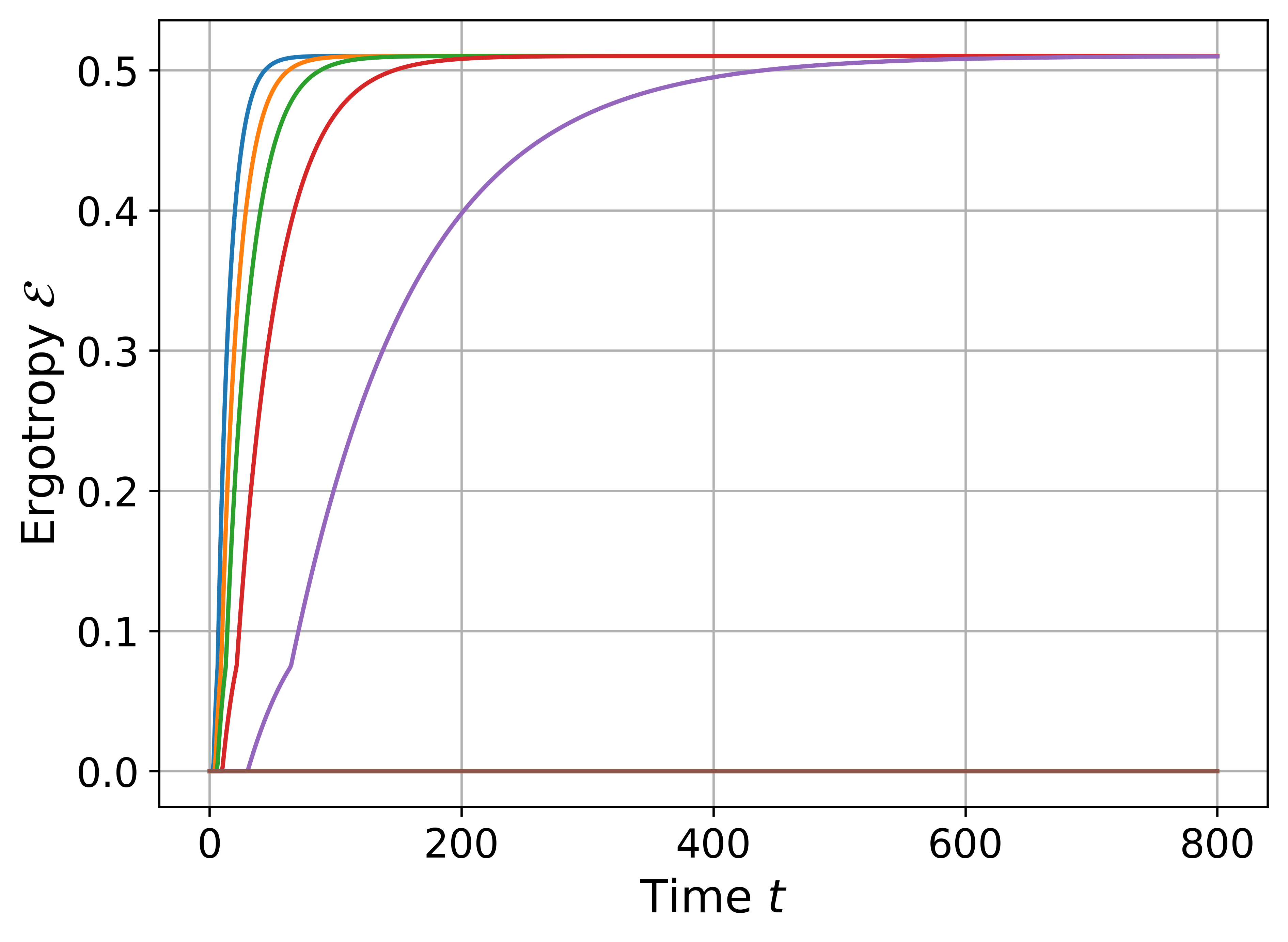}
    \caption{Ergotropy $\mathcal{E}$ as function of time for $\alpha = 0$ (blue curve), $\alpha = 0.3$ (orange curve), $\alpha = 0.5$ (green curve), $\alpha = 0.7$ (red curve), $\alpha = 0.9$ (purple curve) and $\alpha = 1$ (brown curve) for (a) $\alpha^{(-)} = \alpha^{(z)} = 0$ and $\beta = 0.2$; (b) $\alpha^{(-)} = \alpha^{(z)} = 0$ and $\beta = 5$; (c) $\alpha^{(-)} = \alpha^{(z)} = 1$ and $\beta = 0.2$. It can be observed that dephasing mainly slows down the dynamics without changing the final steady-state \textcolor{black}{value of the} ergotropy, with the exception of the case $\alpha = 1$}.
    \label{fig_app:interpolation_channels}
\end{figure*}

We finally fix both $\alpha^{(-)}$ and $\alpha^{(z)}$ to either 0 (fully parallel) or 1 (fully collective) and we interpolate $\alpha$ assigning a different weight on each channel. For each scenario we report the behavior of the lowest and highest values of $\beta$ we fixed in the main text. The results are shown in the three panels of Fig. \ref{fig_app:interpolation_channels} (note that the fully collective case for high values of $\beta$, more precisely for $\beta > \beta_c$, would give zero, or almost zero, ergotropy for all values of $\alpha$. This is the reason why we do not show the relative plot in this figure). Interestingly we notice that increasing the weight of the dephasing channel does not modify the \textcolor{black}{value of the} steady-state ergotropy, but only the time needed to reach it. So, we can say that dephasing does not change the steady-state physics, but the smaller the weight of the dissipative channel, the
longer it takes to reach that value. Conversely, the transient dynamics are more strongly affected by dephasing, as it is shown in panel (b) of Fig. \ref{fig_app:interpolation_channels}. 

\section{Derivation of Eq. (\ref{Passive_Non_Passive_Boundary})} \label{App:beta_c}
Starting from the system of equations reported in Eqs. \eqref{ODEs_twoqubits_collective}, we can obtain a solution for $s(t) \equiv p_{eg}(t) + p_{ge}(t)$ and $c(t)$
\begin{equation}
\begin{cases}
    s(t) = 2\gamma p_{ee}(0) ~t e^{-2 \gamma t} + \frac{s(0)}{2}(1 + e^{-2\gamma t}) + c(0) (e^{-2\gamma t} - 1) \\
    c(t) = \gamma p_{ee}(0)~ t e^{-2 \gamma t} + \frac{s(0)}{4}(e^{-2\gamma t} - 1) + \frac{c(0)}{2} (e^{-2\gamma t} + 1).
\end{cases} \label{s_c_collective}
\end{equation}
Since our initial state is thermal, the initial conditions are
\begin{equation}
    s(0) = \frac{2\cosh(2\beta)}{Z}, \quad c(0) = -\frac{\sinh(2\beta)}{Z},
\end{equation}
with $Z = 2[\cosh(2\beta) + \cosh(2\beta h)]$. From Eq.~\eqref{s_c_collective} we can notice how the steady state depends on the initial conditions, since
\begin{equation} \label{s_c_infty}
\begin{aligned}
    &s(t \to \infty) \equiv s_\infty = \frac{s(0)}{2} - c(0) = \frac{e^{2\beta}}{Z}, \\ &c(t \to \infty) \equiv c_\infty = - \frac{s_\infty}{2}.
\end{aligned}
\end{equation}
\textcolor{black}{To gain a physical understanding of the steady-state behavior, 
it is useful to introduce the symmetric and antisymmetric 
one-excitation states
\begin{equation}
    |\psi^\pm\rangle =
    \frac{|eg\rangle \pm |ge\rangle}{\sqrt{2}}.
\end{equation}
In this basis, the collective jump operator acts as
\begin{equation}
    \begin{aligned}
        &L|ee\rangle = \sqrt{2\gamma}|\psi^+\rangle,\\
        &L|\psi^+\rangle = \sqrt{2\gamma}|gg\rangle,\\
        &L|\psi^-\rangle = 0,\\
        &L|gg\rangle = 0.
    \end{aligned}
\end{equation}
So, the collective dissipative dynamics connects the bright states through the decay sequence
$|ee\rangle \rightarrow |\psi^+\rangle \rightarrow |gg\rangle$, while the singlet state $|\psi^-\rangle$ is dark with respect
to the dissipative channel. Since the latter is also an eigenstate of the XX Hamiltonian, its population is conserved during the evolution. More explicitly, the populations of the symmetric and antisymmetric states are respectively
\begin{equation}
p_\pm(t)=\frac{s(t)}{2}\pm c(t),
\end{equation}
where we have used the fact that $p_{eg}(t)=p_{ge}(t)$ and $c(t)$ is real for the initial conditions considered here. In the long-time limit, one finds both $p_{ee}(\infty)=0$, as follows from the fourth equation of Eq.~\eqref{ODEs_twoqubits_collective}, and $p_+(\infty)=0$. Conversely, the singlet component is preserved, with $p_-(\infty)=p_-(0)=s_\infty$, as we expect from a dark eigenstate. This means that all the initial weight outside the singlet sector eventually accumulates in $|gg\rangle$, while the singlet retains its initial weight. For the thermal initial state considered here, we explicitly have
\begin{equation}
    p_-(\infty)=p_-(0)=
    \langle\psi^-|\rho_\beta|\psi^-\rangle
    =\frac{e^{2\beta}}{Z} = s_{\infty},
\end{equation}
and
\begin{equation}
    p_{gg}(\infty)=1-s_\infty,
\end{equation}
so that the steady state can be written as
\begin{equation}
    \rho_{\mathrm{SS}} =
    (1-s_\infty)|gg\rangle\langle gg|
    +s_\infty|\psi^-\rangle\langle\psi^-|.
\end{equation}
For $h<J$, the steady state is passive if and only if the population of the singlet is greater than or equal to that of $|gg\rangle$, namely
\begin{equation}
    s_\infty\geq 1-s_\infty,
\end{equation}
or equivalently $s_\infty\geq 1/2$. Using the expression for $s_\infty$ in Eq.~\eqref{s_c_infty}, this condition becomes
\begin{equation} \label{Inequality_Beta_Critico}
    \sinh(2\beta)\geq\cosh(2\beta h).
\end{equation}
The critical inverse temperature $\beta_c$ is thus obtained by saturating this inequality, leading to Eq.~\eqref{Passive_Non_Passive_Boundary} of the main text.}

\section{Crossing of ergotropy curves} \label{App:Four_Qubits_Crossings}

\begin{figure}[H]
  \centering

  \begin{subfigure}{\columnwidth}
    \centering
    \includegraphics[width=\columnwidth]{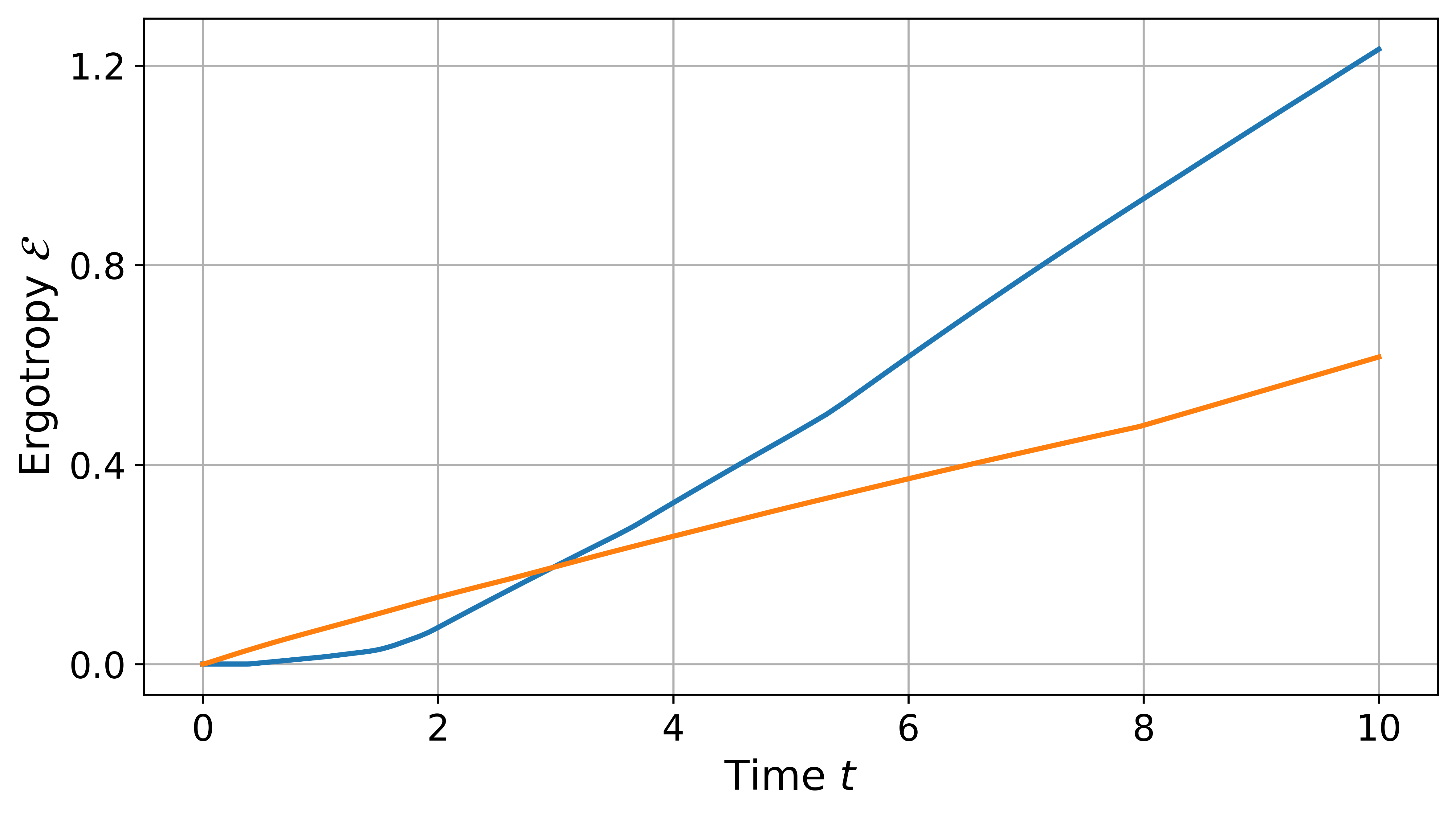}
    \caption{}
   \end{subfigure}

  \vspace{0.4cm}

  \begin{subfigure}{\columnwidth}
    \centering
    \includegraphics[width=\columnwidth]{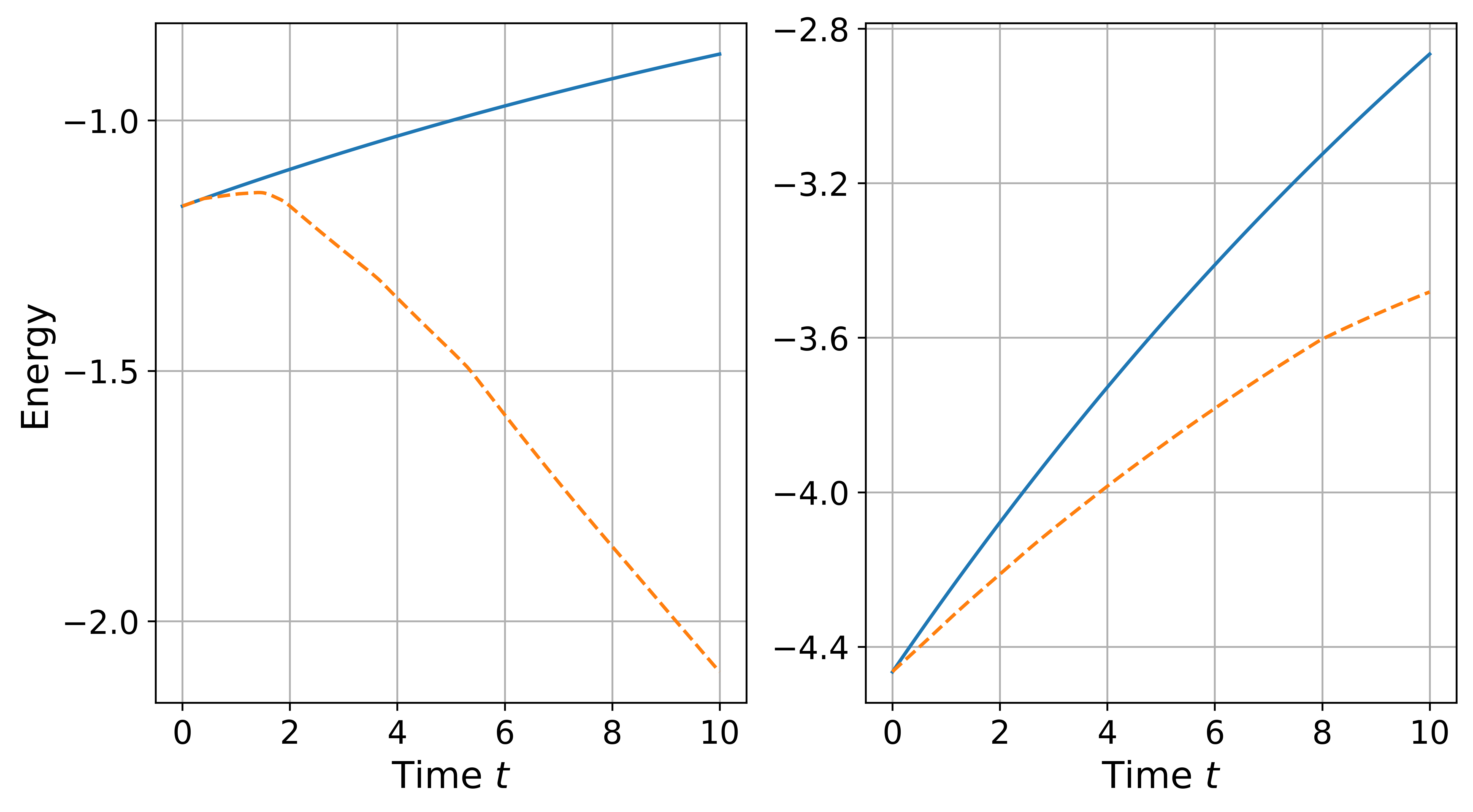}
    \caption{}
  \end{subfigure}
   
   \vspace{0.4cm}

  \begin{subfigure}{\columnwidth}
    \centering
    \includegraphics[width=\columnwidth]{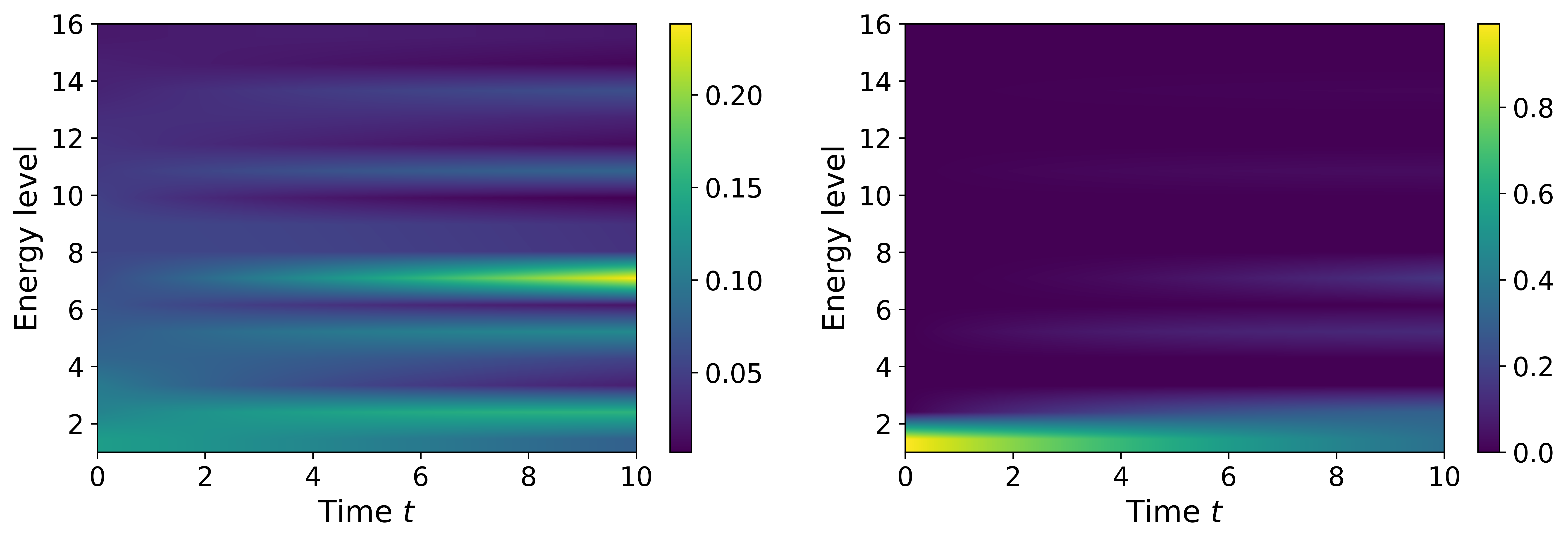}
    \caption{}
  \end{subfigure}

  \caption{(a) Ergotropy as function of time for an initial thermal state with $\beta = 0.2$ (blue curve) and $\beta = 5$ (orange curve). (b) Average energy (blue curve) and passive state's energy (dotted orange curve) as functions of time for $\beta = 0.2$ (left panel) and $\beta = 5$ (right panel). (c) Heatmaps of populations for each energy level with $\beta = 0.2$ (left panel) and $\beta = 5$ (right panel). All plots have been obtained fixing $N = 4$, $h = 0.1$ and $\gamma = 0.05$. The ergotropic Mpemba-like crossing is traced back to eigenvalue crossings that induce sudden \textcolor{black}{changes in the slope of} the passive-state energy.}
  \label{fig_app:two_curves}
\end{figure}

Here, we present a more detailed analysis of the ergotropy in the four-qubit system dissipating through the parallel scheme, focusing on the $\beta = 0.2$ and $\beta = 5$ curves shown in Fig. \ref{fig:four_qubits_parallel} of the main text. In panel (a) of Fig. \ref{fig_app:two_curves}, the crossing between the higher-temperature (blue curve) and lower-temperature (orange curve) curves is clearly visible. \textcolor{black}{Differently from what happens for the two-qubit system, here} the larger Hilbert-space dimension results in a greater number of eigenvalues of the system's density matrix, which in turn allows for multiple crossings in the energy spectrum. \textcolor{black}{This reflects in the fact that the ergotropy curves shown in panel (a) of Fig. \ref{fig_app:two_curves} exhibits several changes in curvature in correspondence} with these spectral crossings. The time at which these crossings occur depends on the temperature of the initial state, with hotter initial states exhibiting earlier crossings than colder ones. After each crossing, the corresponding ergotropy curve becomes steeper, leading to a reordering of the relaxation dynamics.  
In panel (b) of Fig. \ref{fig_app:two_curves} it is possible to observe the implications that such eigenvalues crossings have \textcolor{black}{at the level of the energy}. Here, we compare  for $\beta = 0.2$ (left panel) and $\beta = 5$ (right panel), the average energy $\mathrm{Tr}[\rho(t)H]$ (blue curve) and the passive-state energy $\mathrm{Tr}[\rho_{\text{passive}}(t)H]$ (dotted orange curve), so that the ergotropy at each time corresponds to the difference between \textcolor{black}{these} two quantities. While the average energy in this time interval evolves monotonically, the passive-state energy exhibits sudden \textcolor{black}{changes of slope} occurring after each eigenvalue crossing in the energy spectrum. \textcolor{black}{This behavior is more evident} at lower $\beta$, since higher temperatures lead to a broader population of excited states (as shown in the population heatmaps reported in panel (c) of Fig. \ref{fig_app:two_curves}).
For $\beta = 5$, the eigenvalue crossings of $\rho(t)$ occur later in time, so the corresponding passive-state energy remains comparatively higher during the early stages of evolution. As a result, the ergotropy of the colder initial state initially surpasses that of the hotter one, but once the crossings take place, the situation is inverted, leading to an ergotropic Mpemba-like crossing \textcolor{black}{similar to the one investigated in Ref. \cite{Medina25}}.

\begin{figure}[H]
  \centering

  \begin{subfigure}{\columnwidth}
    \centering
    \includegraphics[width=0.9\columnwidth]{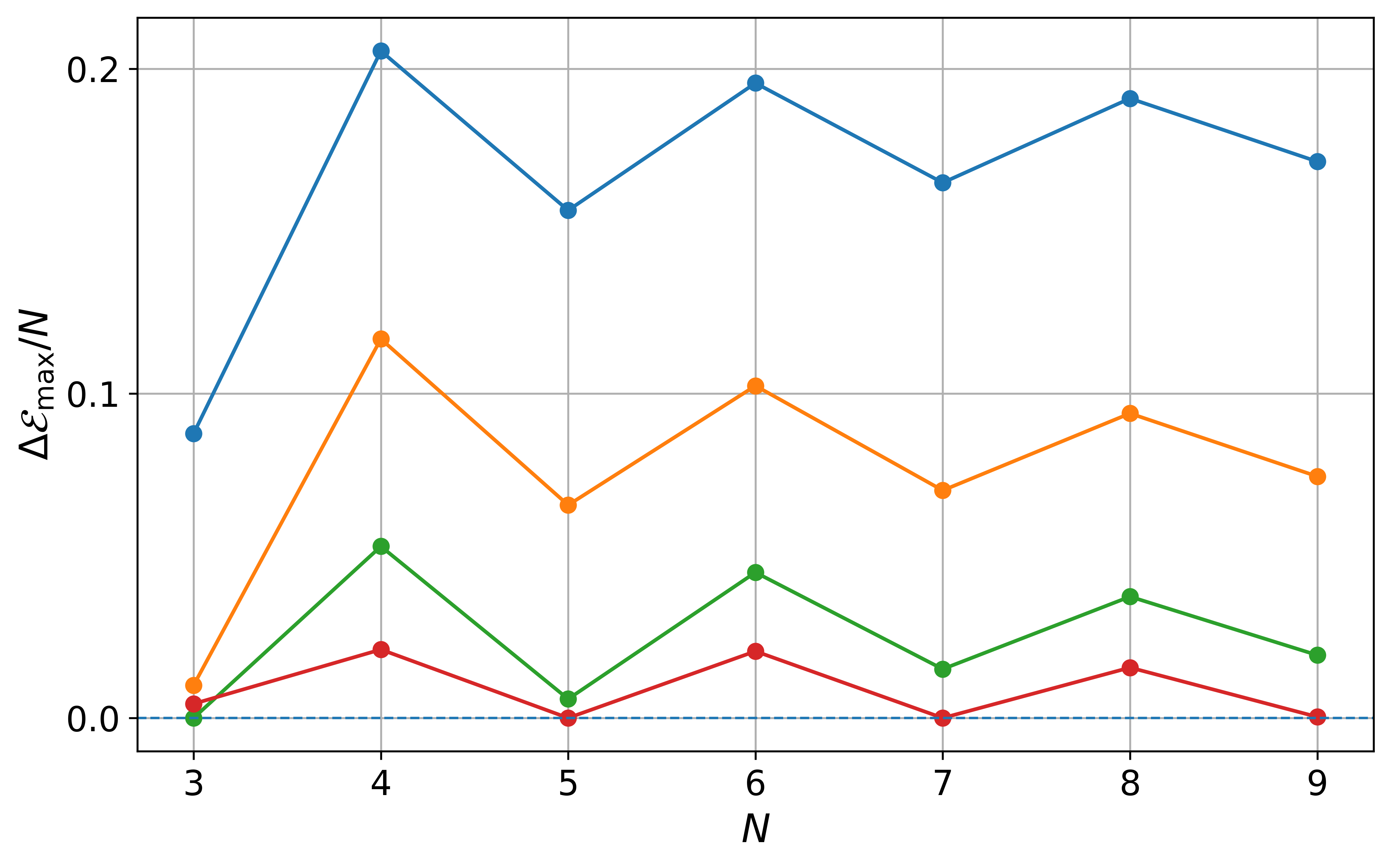}
    \caption{}
    \label{fig:mpemba_strength}
  \end{subfigure}

  \vspace{0.4cm}

  \begin{subfigure}{\columnwidth}
    \centering
    \includegraphics[width=0.9\columnwidth]{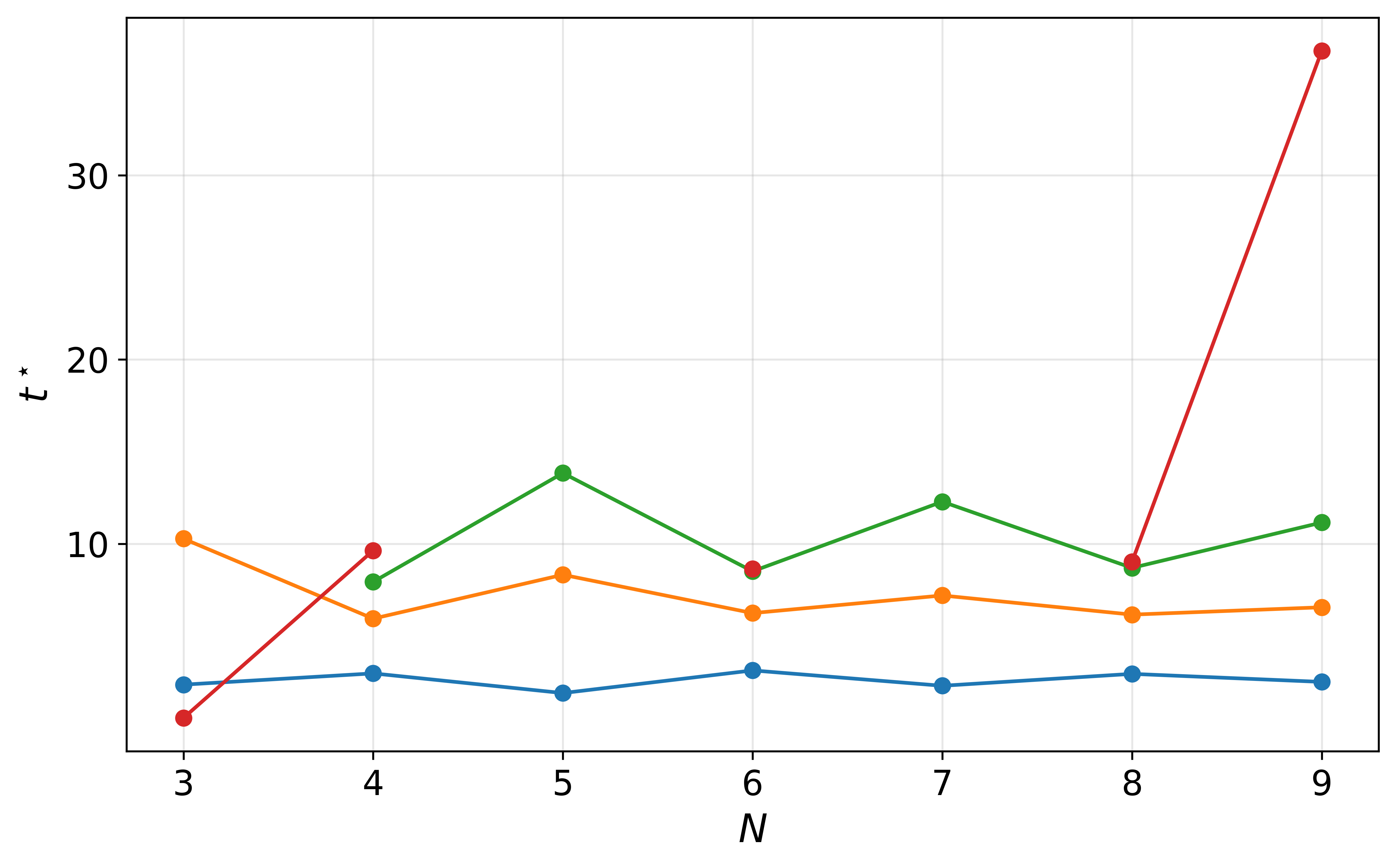}
    \caption{}
    \label{fig:cross_time}
  \end{subfigure}

  \caption{(a) Maximum ergotropic advantage per site at temperature $\beta$ with respect to $\beta = 5$ as a function of $N$. (b) Corresponding first crossing time $t^*$ as a function of $N$. In both panels, the curves refer to $\beta = 0.2$ (blue), $\beta = 0.5$ (orange), $\beta = 1$ (green) and $\beta = 2$ (red), while $N$ is taken in the range $[3,9]$.}
  \label{fig_app:mpemba_approfondito}
\end{figure}

\textcolor{black}{\section{Ergotropic Mpemba-like effect for chain sizes $N \in [3,9]$}\label{App:Mpemba_39} }
\textcolor{black}{To further investigate the dependence of the ergotropic Mpemba-like effect on the system size, we extend the numerical analysis to chains containing up to nine qubits. We consider the same Hamiltonian and local amplitude-damping dynamics discussed in the main text, with $J=1$, $h=0.1$, and $\gamma=0.05$. The initial states are thermal states at inverse temperatures
$\beta\in\{0.2,0.5,1.0,2.0,5.0\}$, and the dynamics is simulated over the time interval $t\in[0,100]$, using 600 equally spaced time points. For each chain length, we take the evolution starting from the coldest initial state, $\beta_{\mathrm{ref}}=5$, as a reference and introduce the ergotropy difference
\begin{equation}
    \Delta\mathcal{E}_{\beta}(t;N)
    =
    \mathcal{E}_{\beta}(t;N)
    -
    \mathcal{E}_{\beta_{\mathrm{ref}}}(t;N),
\end{equation}
as in the main text. A positive value of this quantity indicates that the ergotropy associated with the initial state at inverse temperature $\beta$ exceeds that obtained from the colder initial state. To quantify the magnitude of the effect, we further consider the maximum positive ergotropic advantage,
\begin{equation}
    \Delta\mathcal{E}_{\max}(N,\beta)
    =
    \max_{t\in[0,100]}
    \left\{
    0,\Delta\mathcal{E}_{\beta}(t;N)
    \right\}.
\end{equation}
Moreover, we interpret the crossing times by using the fact that, under local amplitude damping, the energy expectation value of an initially thermal state at inverse temperature $\beta$ obeys
\begin{equation}
E_\beta(t)=-hN+e^{-\gamma t}
\left[E_\beta(0)+hN\right],
\end{equation}
where $E_\beta(t)=\mathrm{Tr}[H\rho_\beta(t)]$. Consequently, the energy difference with respect to the reference state at $\beta_{\mathrm{ref}}=5$ decays exponentially as
\begin{equation}
\Delta E_\beta(t)
\equiv E_\beta(t)-E_5(t)
=e^{-\gamma t}\Delta E_\beta(0).
\end{equation}
Since the ergotropy is given by the difference between the average energy and the passive one, the crossing condition
$\mathcal{E}_\beta(t^*)=\mathcal{E}_5(t^*)$ at a certain time instant $t^*$ can be equivalently expressed as
\begin{equation}
\Delta E_{\mathrm{pas},\beta}(t^*)
=e^{-\gamma t^*}\Delta E_\beta(0),
\end{equation}
where $\Delta E_{\mathrm{pas},\beta}(t)=E_{\mathrm{pas},\beta}(t)-E_{\mathrm{pas},5}(t)$. Thus, the crossing time $t^*$ for each $\beta \neq 5$ is numerically estimated by the intersection between the exponentially decaying difference in average energies and the time-dependent difference in passive energies. The resulting values of $\Delta\mathcal{E}_{\max}(N,\beta)/N$ and the crossing time $t^*$ as functions of $N \in [3,9]$ are reported in Figures \ref{fig:mpemba_strength} and \ref{fig:cross_time} respectively. We observe that, for every system size considered, crossings are consistently found for $\beta=0.2$ and $0.5$, throughout the entire range of system sizes investigated. Interestingly, a clear even-odd modulation of the maximum ergotropic advantage per site appears. In particular, for each value of $\beta$ considered, the values obtained for even chain lengths are larger than those for the neighboring odd chain lengths, with the special cases $\beta = 1$, which shows no crossings for $N = 3$, and $\beta=2$ that shows no crossings for $N=5,7$, with a very weak one for $N = 9$. This behavior highlights the presence of nontrivial finite-size effects in the ergotropic response of the chain.}

\bibliography{references}
\end{document}